\documentclass[12pt,a4paper]{article}

\usepackage{jheppub}

\usepackage{etoolbox}
    \makeatletter
    \patchcmd{\maketitle}{\@fpheader}{}{}{}
    \makeatother

\usepackage[utf8]{inputenc}
\usepackage{lmodern}
\usepackage{exscale}

\usepackage{amsmath}
\usepackage{amssymb}
\usepackage{mathtools}

\usepackage{mathrsfs}

\usepackage{tensor}

\usepackage[low-sup]{subdepth}

\usepackage{graphicx}

\usepackage{xcolor}
\hypersetup{
    colorlinks,
    linkcolor={black},
    citecolor={black},
    urlcolor={black}
}

\usepackage[normalem]{ulem}

\usepackage{cancel}

\title{\boldmath Hamiltonian analysis of \\ 
$\qquad$ metric-affine-$R^2$ theory}

\author[a]{Dra\v{z}en Glavan,}
\emailAdd{glavan@fzu.cz}
\author[b]{Tom Zlosnik,}
\emailAdd{thomas.zlosnik@ug.edu.pl}
\author[c]{Chunshan Lin}
\emailAdd{chunshan.lin@uj.edu.pl}

\affiliation[a]{CEICO, FZU --- 
Institute of Physics of the Czech Academy of Sciences,
	\\
	Na Slovance 1999/2, 182 21 Prague 8, Czech Republic}
\affiliation[b]{Institute of Theoretical Physics and Astrophysics, 
University of Gda\'{n}sk,\\
ul. Wita Stwosza 57, 80-308 Gda\'{n}sk, Poland}
\affiliation[c]{Faculty of Physics, Astronomy and Applied Computer Science, Jagiellonian University, \\
ul. prof. Stanis{\l}awa {\L}ojasiewicza 11, 30-348 Krak\'{o}w, Poland}

\abstract{
Determining the number of propagating degrees of freedom in 
metric-affine theories of gravity requires the use of Hamiltonian
constraint analysis, except in some subclasses of theories. We
develop the technicalities necessary for such analyses and
apply them to the Weyl-invariant and projective-invariant 
case of metric-affine-$R^2$ theory that is known to propagate 
just the graviton. This serves as a check of the formalism
and a case study where we introduce appropriate ADM variables 
for the distortion 3-tensor tensor and its time derivatives,
that will be useful when analyzing more general metric-affine 
theories where the physical spectrum is not known.
}

\begin{document}

\notoc
\maketitle
\titlepage

\section{Introduction}
\label{sec: Introduction}

Theories of gravity are usually formulated in terms of the 
metric~$g_{\mu\nu}$ that describes the geometry of the spacetime
manifold. Accordingly, the dynamics of spacetime is given by
the Einstein equation or generalizations thereof, that is an
equation of motion for the metric. This is the metric 
formulation of gravity that presumes the parallel transport on 
the spacetime manifold is given by the symmetric and 
metric-compatible Levi-Civita connection. However, in general
the connection~$\Gamma^\alpha_{\mu\nu}$ is an independent 
geometrical object. For that reason it is only natural
to consider the possibility that connection is an independent
dynamical field.
The independent connection is often split into
its torsion part that accounts for the anti-symmetric 
part,~$\Gamma^\alpha_{[\mu\nu]} \!\neq\!0$, and non-metricity
that accounts for departure from metric compatibility of the
covariant derivative,~$D^\alpha g_{\mu\nu} \!\neq\!0$.
Such theories of gravity where the spacetime
manifold is described by both~$g_{\mu\nu}$ 
and~$\Gamma^{\alpha}_{\mu\nu}$ are known as metric-affine 
formulations of gravity~\cite{Hehl:1994ue}, and sometimes as
Palatini formulations~(see~\cite{Ferraris:1982} for an account of 
the early days). 

The immediate question when considering the metric-affine 
formulation of gravity is the one of equations of motion
for the independent connection. These are usually concisely
encoded by specifying the action from which the equations of motion
follow by variation with respect to independent dynamical fields.
Even though some actions in the metric formulation and 
the metric-affine formulation might look the same from the
perspective of curvature invariants,
in general such actions describe physically different 
theories~\cite{Exirifard:2007da,Iglesias:2007nv,Borunda:2008kf}. The most well-known example 
happens to be misleading in this regard. That is the one of 
Einstein-Hilbert action written in the metric-affine formulation,
which leads to exactly the same Einstein equations
as the metric formulation,
since the connection is forced to equal the Levi-Civita one
on-shell. This was also demonstrated by Hamiltonian constraint 
analysis~\cite{Kiriushcheva:2010pia,Dadhich:2010xa,Escalante:2013rca}. 
The equivalence of the two
formulations in this case is a coincidence. 
Arguably the simplest generalization to~$f(R)$ theories
of gravity already demonstrates this point clearly.
While the metric formulation of~$f(R)$ theories is known to
propagate three degrees of freedom (see
e.g.~\cite{DeFelice:2010aj,Nojiri:2010wj} for a reviews)
--- the graviton and a scalar
field often referred to as the scalaron --- this is not the
case for the metric-affine counterpart.
Metric-affine~$f(R)$ theories are known to propagate just the 
graviton, and are in fact equivalent to the Einstein-Hilbert 
theory with a cosmological constant~\cite{Ferraris:1992dx,Sotiriou:2009xt}.
in the absence of coupling to matter.

The lesson from the~$f(R)$ example demonstrates the importance 
of understanding what is the physical content
of metric-affine theories, and what is the number and character of
physical degrees of freedom that are propagated by such theories.
Our work is primarily motivated by these questions.
While it is currently not feasible to give an answer in full generality,
considerable progress has been made in this direction. 
There are a number related formulations of 
gravity~\cite{BeltranJimenez:2019esp,Bahamonde:2021gfp}
that are somewhat more restricted, and 
include either torsion or non-metricity 
(or even further restrictions such as Weyl geometry), 
and differ by how these are coupled to matter. 
In some instances determining the number of propagating
degrees of freedom is possible by clever 
field redefinitions/frame transformations, as was possible
for~$f(R)$ theories (by procedure that we recall in 
Sec.~\ref{sec: Equivalence to Einstein-Hilbert}), 
while other instances require the use of Hamiltonian constraint analysis.
Considerable progress has been made in understanding the physical
content of theories generalizing~$f(R)$ to depend on the Ricci tensor 
as well, either with or without 
torsion~\cite{Obukhov:1996ka,Sotiriou:2008rp,Olmo:2009xy,Vitagliano:2010pq,Vitagliano:2010sr,Vitagliano:2013rna,BeltranJimenez:2014iie,BeltranJimenez:2019acz,BeltranJimenez:2020sqf,BeltranJimenez:2020guo,Bejarano:2019zco}
(the latter often referred to as the Palatini formulation).
These studies show that in general metric-affine theories can 
exhibit a different number of gravitational propagating degrees of 
freedom: in some instances theories are similar to~$f(R)$
theories and propagate only a massless graviton, while in
some instances there is also a massive vector 
or other fields that appear in the spectrum.
Some of these additional fields can also be pathological.
Hamiltonian analysis of constraints has been performed for the so-called Poincar\'{e} gauge 
theories~\cite{Blagojevic:2000xd,Yo:1999ex,Yo:2001sy,Blagojevic:2018dpz,Lin:2018awc,Lin:2019ugq}, which
are related to metric-affine theories without non-metricity.
Progress has recently been made towards automating this 
full nonlinear analysis of such theories using computational 
algebra~\cite{Barker:2022kdk}.
Hamiltonian analysis has also been applied to 
so-called teleparallel theories that do not contain the metric,
but are either built only out of torsion~\cite{Blagojevic:2000qs,Li:2011rn,Ferraro:2018tpu,Blagojevic:2020dyq} (dubbed~$f(T)$ theories), or only out of non-metricity
\cite{Hu:2022anq,DAmbrosio:2023asf,Tomonari:2023wcs,Guzman:2023oyl,Gomes:2023tur}~(dubbed~$f(Q)$ theories).

A pertinent question when analyzing the degrees of freedom
is whether the theory exhibits Ostrogradsky instabilities. This
question seems intimately tied to the so-called projective 
symmetry~\cite{Julia:1998ys}, which is a local symmetry under 
particular shift of the independent connection. 
This symmetry seems to discriminate between theories that do not show instabilities 
compared to the ones that do~\cite{Afonso:2017bxr,BeltranJimenez:2019acz,Aoki:2019rvi,Percacci:2020ddy}.
That is why we are inclined to consider theories exhibiting 
projective symmetry. Furthermore, in particle physics and 
quantum gravity Weyl invariance --- the invariance under local rescaling
of the metric --- plays a prominent role.
However, this symmetry cannot be implemented purely geometrically 
in the metric formulation without introducing Ostrogradsky 
instabilities~\cite{Stelle:1977ry,Stelle:1976gc},
but rather requires the introduction of other fields. On the other hand it
is relatively straightforward to implement Weyl invariance in the 
metric-affine formulation of gravity, and moreover it 
provides insight into the renormalizability issues and
opens up the possibility of
generating the Planck scale via symmetry-breaking mechanisms~\cite{Ghilencea:2020piz,Ghilencea:2022lcl,Baldazzi:2021kaf,Melichev:2023lwj,Alvarez:2017spt,Ferreira:2018itt,Olmo:2022ops}. 
We take the approach of considering a purely metric-affine 
Weyl-invariant gravity, where the dynamical content
is the metric $g_{\mu\nu}$ and the connection $\Gamma^{\alpha}_{\mu\nu}$, and do not introduce additional fields such a compensating scalar field (dilaton).
What we would like to know is how many and what kind of degrees of freedom do such theories propagate.

For that reason we are motivated to consider a general class
of metric-affine gravity theories that are projective-invariant,
Weyl-invariant, and whose action is built out of invariants
composed of contracted Riemann tensors (we give the list
of such invariants by the end of Sec.~\ref{sec: Metric-affine (Palatini) formulation}). 
We would like to know how many
degrees of freedom do such theories propagate. Determining this requires
the use of Hamiltonian analysis of constraints.
However, undertaking this task will be rather involved,
evident already from the linearized level~\cite{Percacci:2020ddy}.
For that reason, this paper is devoted to setting up the formalism
for such an analysis by introducing appropriate Arnowitt-Deser-Misner 
(ADM) variables for the 
problem, and applying the formalism to the relatively simple case of 
metric-affine~$R^2$ gravity. 
This allows us to set up the formalism and provide a check for it. 
Our analysis is closely related to the one reported 
in~\cite{Olmo:2011fh}, that reported a Hamiltonian analysis of
a particular Brans-Dicke theory~\cite{Wang:2004pq,Olmo:2005zr,Sotiriou:2006hs}
that is equivalent to metric-affine~$R^2$.
We also point out the importance of the last step in the analysis,
which is the reduction of phase space by solving the second-class constraints.
From the technical side, this allows to understand how a 
complicated-looking theory could be a simpler theory in disguise,
and on the conceptual side it provides insight how some theories 
which ostensibly describe the coupling of a certain type of matter to gravity
might have a purely geometrical origin and 
interpretation~\cite{Buchdahl:1979ut,BeltranJimenez:2014iie,Barker:2023fem,Vitagliano:2010pq}.
For the particular example of metric-affine~$R^2$ theory at hand
it is rather interesting that the Weyl-invariant theory that contains 
no scales is at the end of
the day equivalent to the Einstein-Hilbert theory with a cosmological 
constant, where both the Newton's constant and the cosmological constant
are related to an arbitrary scale introduced by hand.
In that sense the theory can be seen as a Weyl-invariant
geometrization of the Einstein-Hilbert theory.
It can also be seen as an extension of the Weyl geometry to
a more general metric-affine structure~\cite{Delhom:2019yeo}.

The structure of the paper is as follows: In 
Section~\ref{sec: Metric-affine (Palatini) formulation} we summarize the mathematical structure of metric-affine gravity and the construction 
of projective- and Weyl-invariant actions. In 
Section~\ref{sec: Equivalence to Einstein-Hilbert} we 
recall how field redefinitions in the Lagrangian formulation of
metric-affine~$R^2$ theory are used to demonstrate its equivalence to 
the Einstein-Hilbert theory in the presence of a cosmological constant. 
In Section~\ref{sec: ADM decomposition} we present the 
ADM decomposition of the metric-affine 
gravitational fields and their time derivatives,
appropriate for any metric-affine theory, and we apply it to the~$R^2$
model. The following Section~\ref{sec: Canonical action} is devoted to
deriving the canonical action for this model. In Section~\ref{sec: Constraint analysis} we present the analysis of the propagation of constraints in 
the canonical formulation of the~$R^2$ theory and characterize and 
determine the number of physical propagating degrees of freedom. 
In Section~\ref{sec: Reducing phase space} we perform a reduction of 
the phase space by solving the secondary constraints, 
firstly showing the equivalence of the theory to a particular Brans-Dicke 
theory, and secondly, via canonical transformation showing equivalence 
to the Hamiltonian formulation of General Relativity in the presence of 
a cosmological constant. In Section~\ref{sec: Discussion} we present our 
conclusions.

\section{Metric-affine formulation and local symmetries}
\label{sec: Metric-affine (Palatini) formulation}

The metric-affine formulation of gravity, also known as the Palatini formulation,
utilizes two independent geometric objects --- the metric tensor~$g_{\mu\nu}$,
and the independent affine connection~${\Gamma^{\alpha}}_{\mu\nu}$. Unlike in 
the standard metric formulation of gravity, the connection is not a priori assumed
to be symmetric in the indices,~${\Gamma^\alpha}_{[\mu\nu]} \!\neq\! 0$,
nor is it assumed to be metric-compatible,~$D_\alpha g_{\mu\nu} \!\neq\! 0$, with~$D_\alpha$
standing for the covariant derivative defined with respect to the affine connection.
The curvature is described by the Riemann tensor,
\begin{equation}
{R^\sigma}_{\rho\mu\nu}
	= \partial_\mu {\Gamma^{\sigma}}_{\nu\rho}
	- \partial_\nu {\Gamma^{\sigma}}_{\mu\rho}
	+ {\Gamma^{\alpha}}_{\nu \rho} {\Gamma^{\sigma}}_{\mu\alpha}
	- {\Gamma^{\alpha}}_{\mu\rho} {\Gamma^{\sigma}}_{\nu\alpha} \, ,
\label{Riemann tensor}
\end{equation}
where we adopt the conventions from a brief review on metric-affine gravity 
theories~\cite{Jarv:2018bgs}, so that Greek indices run over 4 space-time dimensions.
There are three independent contractions of the Riemann tensor
yielding 2-tensors,
\begin{equation}
R_{\mu \nu} = {R^\alpha}_{\mu\alpha\nu} \, ,
\qquad \qquad
\overline{R}_{\mu\nu} = {R^\alpha}_{\alpha\mu\nu} \, ,
\qquad \qquad
\hat{R}_{\mu\nu} = 
	g_{\mu \alpha} g^{\rho\sigma} {R^{\alpha}}_{\rho\sigma\nu} \, .
\label{Ricci tensors}
\end{equation}
respectively called the Ricci tensor (that is now generally not symmetric), 
the homothetic curvature, and the co-Ricci tensor, while there is still a unique 
Ricci scalar formed by contracting the inverse metric into the Ricci tensor,
\begin{equation}
R = g^{\mu\nu} R_{\mu\nu} \, .
\label{Ricci scalar}
\end{equation}

It is possible to reformulate metric-affine theories as purely metric theories interacting
with a rank-3 tensor field. This is accomplished by a field redefinition of the affine 
connection,
\begin{equation}
{\Gamma^\alpha}_{\mu\nu} = \mathring{\Gamma}^\alpha_{\mu\nu}
	+ { \Xi^{\alpha}}_{\mu\nu} \, ,
\label{connection split}
\end{equation}
that is now shifted by the Levi-Civita connection (the Christoffel symbol), 
\begin{equation}
\mathring{\Gamma}^\alpha_{\mu\nu} =
	\frac{1}{2} g^{\alpha\beta} \Bigl( \partial_\mu g_{\nu\beta}
		+ \partial_\nu g_{\mu\beta}
		- \partial_\beta g_{\mu\nu} \Bigr) \, ,
\label{Christoffel}
\end{equation}
leaving a proper rank-3 tensor~$\Xi_{\alpha\mu\nu}$,
known as the distortion tensor. This field redefinition induces a split of the
Ricci scalar~(\ref{Ricci scalar}),
\begin{equation}
R = \mathring{R} + 
	\mathring{\nabla}_\mu \Bigl( {\Xi^{\mu \nu}}_{\nu} - {\Xi_{\nu}}^{\nu \mu} \Bigr)
		+ {\Xi^{\mu}}_{\mu \alpha} { \Xi^{\alpha \nu}}_{\nu}
		- {\Xi^{\alpha}}_{\mu \nu} { \Xi^{\mu\nu}}_{\alpha} \, ,
\label{R in Xi}
\end{equation}
where~$\mathring{R}$ is the usual metric Ricci scalar,
\begin{equation}
\mathring{R} = g^{\mu\nu} \Bigl(
	\partial_\rho \mathring{\Gamma}^{\rho}_{\nu\mu}
	- \partial_\nu \mathring{\Gamma}^{\rho}_{\rho \mu}
	+ \mathring{\Gamma}^{\rho}_{\rho \sigma} \mathring{\Gamma}^{\sigma}_{\nu \mu}
	- \mathring{\Gamma}^{\rho}_{\nu \sigma} \mathring{\Gamma}^{\sigma}_{\rho \mu}
	\Bigr) \, ,
\end{equation}
and~$\mathring{\nabla}_\mu$ is the covariant derivative with respect to the metric-compatible
Levi-Civita connection~$\mathring{\Gamma}^\alpha_{\mu\nu}$. It is conventional to decompose the 
distortion tensor,
\begin{equation}
\Xi_{\alpha\mu\nu} = K_{\alpha\mu\nu} + L_{\alpha\mu\nu} \, ,
\label{distortion decomposition}
\end{equation}
into the contortion tensor~$K_{\alpha\mu\nu} \!=\! K_{\alpha]\mu[\nu}$,
and the disformation tensor~$L_{\alpha\mu\nu} \!=\! L_{\alpha(\mu\nu)}$,
which are themselves expressed in terms of the torsion 
tensor~$T_{\alpha\mu\nu}\!=\!T_{\alpha[\mu\nu]}$
and the non-metricity tensor~$Q_{\alpha\mu\nu}\!=\!Q_{\alpha(\mu\nu)}$,
\begin{align}
&
K_{\alpha\mu\nu}
	= \frac{1}{2} \Bigl( T_{\mu\alpha\nu} + T_{\nu\alpha\mu} + T_{\alpha\mu\nu} \Bigr) \, ,
\qquad
L_{\alpha\mu\nu}
	= \frac{1}{2} \Bigl( - Q_{\mu\alpha\nu} - Q_{\nu\alpha\mu} + Q_{\alpha\mu\nu} \Bigr) \, .
\label{KL decomposition}
\end{align}
The latter two tensors have a clear physical interpretation,
as when expressed in terms of the independent affine connection,
\begin{align}
T_{\alpha \mu \nu} 
	\equiv{}&
	g_{\alpha \beta} {\Gamma^{\beta}}_{[\mu\nu]}
	= \Xi_{\alpha \mu \nu} - \Xi_{\alpha \nu \mu}
	= T_{\alpha[\mu\nu]}
	\, ,
\\
Q_{\alpha\mu\nu}
	\equiv{}& 
	D_\alpha g_{\mu\nu}
	= 
	- \Xi_{\mu\alpha\nu} 
	- \Xi_{\nu\alpha \mu}
	= 
	Q_{\alpha(\mu\nu)} \, .
\label{nonmetricity tensor}
\end{align}
they represent the departure from the symmetric connection, and 
departure from metric compatibility of the connection. 

\medskip

The theory that we examine in this work falls under a broader
class of the so-called metric-affine~$f(R)$ theories,
that are given by the action,
\begin{equation}
S\bigl[ g_{\mu\nu} , \Xi_{\alpha\mu\nu} \bigr] = 
	\int\! d^{D\!}x \, \sqrt{-g} \, f(R) \, ,
\label{f(R) action}
\end{equation}
where 
the Ricci scalar is given in~(\ref{R in Xi}),
$g \!=\! {\rm det} (g_{\mu\nu})$ is the metric determinant, 
and $f$ is an arbitrary well-behaved function.
It is useful to understand the local symmetries that these theories possess
before embarking on the analysis.

Firstly, the action for any metric-affine theory ought to be invariant
under diffeomorphism transformations, under which the metric and the affine
connection transform infinitesimally as,
\begin{align}
&
g_{\mu\nu} 
    \longrightarrow
    g_{\mu\nu} + \mathring{\nabla}_{(\mu} \zeta_{\nu)} 
    \, ,
\label{diff trans}
\\
&
\Xi_{\alpha\mu\nu}
    \longrightarrow
    \Xi_{\alpha\mu\nu}
    +
    \zeta^\beta \mathring{\nabla}_\beta \Xi_{\alpha\mu\nu}
    + \Xi_{\beta\mu\nu} \mathring{\nabla}_\alpha \zeta^\beta
    + \Xi_{\alpha\beta\nu} \mathring{\nabla}_\mu \zeta^\beta
    + \Xi_{\alpha\mu\beta} \mathring{\nabla}_\nu \zeta^\beta
    \, ,
\nonumber
\end{align}
where~$\zeta_\mu$ is an infinitesimal vector field.

Secondly, the Ricci scalar $R$ is invariant under so-called projective transformations
(see~\cite{Iosifidis:2018zwo} for a good discussion on
the physical interpretation),
under which only the affine connection (or the distortion tensor) transforms,
\begin{equation}
g_{\mu\nu} \longrightarrow g_{\mu\nu} \, ,
\qquad \quad
\Xi_{\alpha \mu \nu} \longrightarrow
	\Xi_{\alpha \mu \nu} + g_{\alpha \nu} \zeta_\mu 
 \, ,
\label{projective transformation}
\end{equation}
where $\zeta_\mu$ is again an arbitrary space-time dependent vector.
Thus the metric-affine-$f(R)$ theories in~(\ref{f(R) action}) inherit this
symmetry.
This is another local symmetry of the theory. This can also be viewed as a 
gauge symmetry of the distortion field.

Thirdly, in addition to diffeomorphism invariance, and projective invariance,
in $D\!=\!4$ spacetime dimensions specifically, and for~$f(R)\!=\!R^2$ 
there is 
another symmetry under local Weyl (conformal) transformations,
\begin{equation}
g_{\mu\nu} \longrightarrow e^{\sigma} g_{\mu\nu} \, ,
\qquad \quad
\Xi_{\alpha \mu \nu} \longrightarrow
	\Xi_{\alpha \mu \nu} 
    - g_{\alpha(\mu} \mathring{\nabla}_{\nu)} \sigma
    + \frac{1}{2} g_{\mu\nu} 
        \mathring{\nabla}_\alpha \sigma
 \, , 
\label{Weyl trans}
\end{equation}
where~$\sigma$ is an arbitrary scalar function.
Since the metric determinant and the Ricci scalar transform as,~\footnote{
The metric Ricci tensor transforms under Weyl rescaling~(\ref{Weyl trans}) of the 
metric as,
\begin{equation*}
\mathring{R} \longrightarrow e^{-\sigma} \biggl[ \mathring{R}
	- 3 \mathring{\nabla}^\mu \mathring{\nabla}_\mu \sigma
	- \frac{3}{2} ( \mathring{\nabla}^\mu \sigma ) 
        ( \mathring{\nabla}_\mu \sigma ) \biggr] 
	\, .
\end{equation*}
}
\begin{equation}
g \longrightarrow e^{D\sigma} g \, ,
\qquad \qquad
R \longrightarrow e^{-\sigma} R  \, ,
\end{equation}
the action remains invariant under such rescaling.
Therefore, this particular case, that we examine in this work, should be treated with special
care since it ought to exhibit more first-class constraints compared to the rest of the
theories in~(\ref{f(R) action}).

\medskip

Given the projective invariance~(\ref{projective transformation}) of the theory
we study, we find it more convenient to introduce somewhat non-standard tensor
variables compared to the ones introduced in the decomposition of the distortion tensor 
in~(\ref{distortion decomposition}) and~(\ref{KL decomposition}). While we
still find it convenient to use non-metricity, rather than employing the torsion field
directly it is better to use a projective-invariant 3-tensor,
\begin{equation}
B_{\alpha\mu\nu} \equiv
	\Xi_{\mu\alpha\nu} - \Xi_{\nu\alpha\mu}
	=
	2 K_{\mu\alpha\nu} + Q_{\mu\alpha\nu} - Q_{\nu\alpha\mu}
	= B_{\alpha[\mu\nu]} \, ,
\label{B tensor}
\end{equation}
that is a linear combination of (con-)torsion and non-metricity.
We henceforth refer to this tensor simply as the~$B$-tensor, for the lack of a better word.
Expressing the Ricci scalar in terms of these two tensor gives,
\begin{equation}
R = \mathring{R}  
	- \mathring{\nabla}_\mu {B_{\nu}}^{\nu \mu} 
	- \frac{1}{4} {B^{\mu}}_{\mu \alpha} { B_{\nu}}^{\nu \alpha } 
	+ \frac{1}{4} B_{\mu \nu \alpha } B^{\nu \mu \alpha}
	+ \frac{1}{4} {Q^{\mu}}_{\mu \alpha} { Q_{\nu}}^{\nu \alpha } 
	- \frac{1}{4} Q_{\mu\nu\alpha} Q^{\nu\mu\alpha}
	\, .
\label{R decomp QB}
\end{equation}
The utility of using the~$B$-tensor in theories with projective symmetry 
is that it is independent of any ambiguities associated with choosing
a gauge for projective symmetry transformations.

\medskip

There are a number of metric-affine theories that are both projective-invariant and
Weyl-invariant, and the~$R^2$ theory that this work is devoted to is just one particular 
instance of it. The most general theory can be constructed by first identifying all the projective-invariant components of the curvature tensors in~(\ref{Riemann tensor}) and~(\ref{Ricci tensors}).
For this we first need to work out how each of them transforms under 
projective transformations,
\begin{align}
&
\qquad
R_{\rho\sigma\mu\nu} 
	\longrightarrow
	R_{\rho \sigma \mu \nu} 
	+ 2 g_{\rho \sigma} \partial_{[\mu} \zeta_{\nu]}
	\, ,
\qquad
R_{\mu\nu} 
	\longrightarrow
	R_{\mu \nu} 
	+ 2 \partial_{[\mu} \zeta_{\nu]} \, ,
\nonumber \\
&
\overline{R}_{\mu\nu} 
	\longrightarrow
	\overline{R}_{\mu \nu}
	+ 2 D \partial_{[\mu} \zeta_{\nu]} \, ,
\qquad
\hat{R}_{\mu\nu} 
	\longrightarrow
	\hat{R}_{\mu \nu} 
	+ 2 \partial_{[\mu} \zeta_{\nu]}
	\, ,
\qquad
R 
	\longrightarrow
R \, .
\label{projective transformations}
\end{align}
which then allows identification of the invariant components,
\begin{equation}
R_{[\rho\sigma]\mu\nu} 
\, ,
\quad \
R_{(\mu\nu)} 
\, ,
\quad \
\hat{R}_{(\mu\nu)} 
\, ,
\quad \
\check{R}_{\mu\nu} \equiv
R_{[\mu\nu]} - \hat{R}_{[\mu\nu]} 
\, ,
\quad \
\widetilde{R}_{\mu\nu} \equiv
\hat{R}_{[\mu\nu]} - \frac{1}{D} \overline{R}_{\mu\nu} 
\, ,
\quad \
R 
\, .
\label{projective invariants}
\end{equation}
In addition, Weyl invariance of the theory in~$D\!=\!4$ will be respected if
we form quadratic invariants from these invariant components,
\begin{align}
&
R^2
\, ,
\qquad
R^{(\mu\nu)} R_{(\mu\nu)}
\, ,
\qquad
\hat{R}^{(\mu\nu)} R_{(\mu\nu)}
\, ,
\qquad
\hat{R}^{(\mu\nu)} \hat{R}_{(\mu\nu)}
\, ,
\qquad
\widetilde{R}_{\mu\nu} \widetilde{R}^{\mu\nu}
\, ,
\qquad
\widetilde{R}_{\mu\nu} \check{R}^{\mu\nu}
\, ,
\nonumber \\
&
\check{R}_{\mu\nu} \check{R}^{\mu\nu}
\, ,
\qquad
R^{[\rho\sigma]\mu\nu} R_{[\rho\sigma]\mu\nu} 
\, ,
\qquad
R^{[\rho\sigma]\mu\nu} R_{[\mu\nu]\rho\sigma}
\, ,
\qquad
R^{[\rho\sigma]\mu\nu} R_{[\rho\mu]\sigma\nu} 
\, .
\label{projective quadratic invariants}
\end{align}
The terms we list above are parity-even, but it is also possible to
consider parity-odd quadratic 
invariants~\cite{Jimenez:2022hcz,Gialamas:2022xtt}.
Therefore, the most general parity-even, Weyl-invariant, and 
projective-invariant theory should be given by a Lagrangian 
that is a linear combination of the quadratic invariants
in~(\ref{projective quadratic invariants})
(though one linear combination is a generalization of the 
Gauss-Bonnet term~\cite{Borunda:2008kf} and can be discarded in~$D\!=\!4$).
Our reason for studying the Hamiltonian constraint analysis of
the simple instance of~$R^2$ theory is to develop the methodology
appropriate for such analyses of the more general Weyl-invariant
and projective-invariant theories outlined above.

\section{Equivalence to Einstein-Hilbert}
\label{sec: Equivalence to Einstein-Hilbert}

Metric-affine theories in general differ from their metric counterparts, 
even if their Lagrangians are written in terms of the same curvature invariants.
It is the variational principle associated to the independent connection
that makes all the difference.
In this section we recall how this works for~$f(R)$ theories
and in particular~$R^2$ theories.
While metric~$f(R)$ theories are known to propagate three degrees 
of freedom,
metric-affine~$f(R)$ theories are in fact equivalent to pure Einstein-Hilbert gravity
with a cosmological constant, and propagate just the graviton.
The simplest way of demonstrating both claims consists of a series of frame 
transformations/field redefinitions.
It will become evident that the Weyl-invariant case~$f(R)\!=\!R^2$
and~$D\!=\!4$ requires special attention due to enhanced local symmetry.
Supplying this proof in the Hamiltonian formalism is the subject matter of 
sections that follow.

The first step of the proof is deriving an on-shell equivalent action 
to~(\ref{f(R) action}) that is better adapted to frame transformations. 
We introduce an auxiliary scalar~$\varphi$ that is equal to the Ricci scalar 
on-shell, which is enforced by a Lagrange multiplier~$\lambda$,
\begin{equation}
S\bigl[ g_{\mu\nu} , \Xi_{\alpha \mu \nu}, \varphi ,\lambda \bigr]
	=
	\int\! d^{D\!} x \, \sqrt{-g} \, \Bigl[ f(\varphi) + \lambda (\varphi \!-\! R)  \Bigr]
	\, .
\label{intermediate S}
\end{equation}
The variation with respect to~$\varphi$ generates an on-shell equation,
\begin{equation}
f'(\varphi) + \lambda \approx 0 \, .
\end{equation}
We are allowed to solve this equation for~$\lambda$ and plug the solution back into 
the action~(\ref{intermediate S}), since the resulting action yields the same on-shell
equations,
\begin{equation}
S\bigl[ g_{\mu\nu} , \Xi_{\alpha \mu \nu}, \varphi \bigr]
	=
	\int\! d^{D\!} x \, \sqrt{-g} \, \Bigl[ f(\varphi) 
    - \varphi  f'(\varphi)
		+ f'(\varphi) R  \Bigr]
	\, .
\label{intermediate S 2}
\end{equation}
Note that this procedure presumes that on shell~$f''(\varphi) \! \not\approx \! 0$.
the proof that metric~$f(R)$ theories propagate a scalaron starts in the
very same way.
However, the big difference here is in the fact that~$R$ depends on the 
distortion tensor,
which is manifested when the Ricci scalar is written out as in~(\ref{R in Xi}),
\begin{align}
\MoveEqLeft[4]
S\bigl[ g_{\mu\nu}, \Xi_{\alpha\mu\nu}, \varphi \bigr]
	=
	\int\! d^{D\!}x \, \sqrt{-g} \, \biggl[
		f(\varphi) - \varphi f'(\varphi) + f'(\varphi) \mathring{R}
\nonumber \\
&	
	- f''(\varphi) \bigl( \mathring{\nabla}_\mu \varphi \bigr)
		\Bigl( {\Xi^{\mu \nu}}_{\nu} - {\Xi_{\nu}}^{\nu \mu} \Bigr)
	+ f'(\varphi) \Bigl( {\Xi^{\mu}}_{\mu \nu} {\Xi^{\nu \alpha}}_{\alpha}
	- {\Xi^{\mu}}_{\nu \alpha} {\Xi^{\nu\alpha}}_{\mu} \Bigr)
		\biggr] \, .
\label{some action}
\end{align}
The first line is just what we would have in the metric~$f(\mathring{R})$ theory,
while the second line is the additional content of the metric-affine formulation.

At this point it is possible to solve the algebraic equations for the 
distortion tensor, and, upon plugging the solution back into the 
action~(\ref{some action}) and introducing~$\sigma\!=\!f'(\varphi)$,
to obtain an equivalent Brans-Dicke 
formulation,~\cite{Wang:2004pq,Olmo:2005zr,Sotiriou:2006hs}
\begin{equation}
S\bigl[ g_{\mu\nu} , \sigma \bigr]
    =
    \int\! d^{D\!}x \, \sqrt{-g} \, 
    \biggl[
    \sigma \mathring{R}
    -
    \frac{\omega}{\sigma}
    \bigl( \mathring{\nabla}^\mu\sigma \bigr)
    \bigl( \mathring{\nabla}_\mu \sigma \bigr)
    -
    V(\sigma)
    \biggr]
    \, ,
\label{Brans-Dicke}
\end{equation}
with~$\omega\!=\! - (D\!-\!1)/(D\!-\!2) \! \xrightarrow{D\to4} -3/2$,
where the potential defined implicitly,~$V(\sigma) \!=\! \varphi f'(\varphi) \!-\! f(\varphi) $.
However, 
the physical content of the theory is better revealed by
conformally rescaling the metric,
\begin{equation}
g_{\mu\nu} \longrightarrow \bigl[ f'(\varphi) \kappa^2 \bigr]^{\frac{-2}{D-2}} g_{\mu\nu} \, ,
\label{g conformal rescaling}
\end{equation}
where~$\kappa$ is some dimensionful parameter that makes the conformal
factor dimensionless,
followed by a compensating rescaling and shifting of the distortion field,
\begin{equation}
\Xi_{\alpha\mu\nu} \longrightarrow
	\bigl[ f'(\varphi) \kappa^2 \bigr]^{\frac{-2}{D-2}} \biggl[ \Xi_{\alpha \mu \nu}
	+ \frac{f''(\varphi)}{(D\!-\!2) f'(\varphi)}
		\Bigl( g_{\alpha \mu} \mathring{\nabla}_\nu\varphi
			+ g_{\alpha \nu} \mathring{\nabla}_\mu\varphi
			- g_{\mu\nu} \mathring{\nabla}_\alpha\varphi \Bigr)
	\biggr] \, .
\label{3tensor frame trans}
\end{equation}
These transformations put the action in the form of the Einstein-Hilbert theory with 
an auxiliary scalar and an auxiliary 3-tensor field,
\begin{align}
\MoveEqLeft[3]
S\bigl[ g_{\mu\nu}, \Xi_{\alpha\mu\nu}, \varphi \bigr] 
=
	\int\! d^{D\!}x \, \sqrt{-g} \, \biggl[
		\frac{\mathring{R}}{\kappa^2}
		+
		\bigl[ f'(\varphi) \kappa^2 \bigr]^{\frac{-D}{D-2}} \Bigl[ f(\varphi) - \varphi f'(\varphi) \Bigr] 
\nonumber \\
&	\hspace{6cm}
	+ \frac{1}{\kappa^2} 
		\Bigl( {\Xi^{\mu}}_{\mu \nu} {\Xi^{\nu \alpha}}_{\alpha}
			- {\Xi^{\mu}}_{\nu \alpha} {\Xi^{\nu\alpha}}_{\mu} \Bigr)
		\biggr] \, .
\end{align}
Note that the compensating 3-tensor transformation in~(\ref{3tensor frame trans})
cancels the would-be kinetic term of the scalar field, which therefore does not propagate.
The 3-tensor either vanishes or is undetermined on-shell, and can be eliminated from the action, while the auxiliary scalar on-shell takes a particular constant value,
\begin{equation}
0 \approx
	\frac{\delta S}{\delta \varphi}
	=
	\sqrt{-g} \frac{\partial}{\partial\varphi} 
		\biggl( \bigl[ f'(\varphi) \kappa^2 \bigr]^{\frac{-D}{D-2}} \Bigl[ f(\varphi) 
			- \varphi f'(\varphi) \Bigr]  \biggr) \biggr|_{\varphi=\varphi_0}  \, ,
\end{equation}
and engenders a (possibly vanishing) cosmological constant,
\begin{equation}
- (D\!-\!2) \Lambda_\kappa
	\equiv
- (D\!-\!2) \kappa^{\frac{-2D}{D-2}} \lambda
	 \equiv \bigl[ f'(\varphi_0) \kappa^2 \bigr]^{\frac{-D}{D-2}} \Bigl[ f(\varphi_0) 
			- \varphi f'(\varphi_0) \Bigr] \, .
\end{equation}
It is permissible to plug in~$\varphi\!\approx\!\varphi_0$ back into the action 
as a strong equality, producing the Einstein-Hilbert action,
\begin{align}
\MoveEqLeft[3]
S[ g_{\mu\nu} ] 
=
	\int\! d^{D\!}x \, \sqrt{-g} \, 
		\biggl[
		\frac{ \mathring{R}- (D\!-\!2) \Lambda_\kappa }{\kappa^2}
		\biggr]
		\, .
\end{align}

\bigskip

A particular case of interest to us
is~$D\!=\!4$ and~$f(R)\!=\!R^2$, which is Weyl-invariant, and thus contains
no scale in the original formulation of the action~(\ref{f(R) action}),
\begin{equation}
S \bigl[ g_{\mu\nu}, \Xi_{\alpha\mu\nu} \bigr]
	= \int\! d^4x \, \sqrt{-g} \, R^2 \, .
\label{Palatini R2}
\end{equation}
It has a very special property that
applying to it the field redefinitions~(\ref{g conformal rescaling})
and~(\ref{3tensor frame trans}) completely removes the auxiliary scalar,
\begin{equation}
S \bigl[ g_{\mu\nu}, \Xi_{\alpha\mu\nu}, \varphi \bigr] =
	\int\! d^{4\!}x \, \sqrt{-g} \, 
		\frac{1}{\kappa^2} 
		\biggl[
		\mathring{R}
		- \frac{1}{4\kappa^2}
		+ {\Xi^{\mu}}_{\mu \nu} {\Xi^{\nu \alpha}}_{\alpha}
		- {\Xi^{\mu}}_{\nu \alpha} {\Xi^{\nu\alpha}}_{\mu} 
		\biggr] \, .
\end{equation}
The distortion field can again be solved for on-shell and simply drops out from 
the action, leaving just the Einstein-Hilbert action with a cosmological constant,
\begin{equation}
S [ g_{\mu\nu} ] =
	\int\! d^{4\!}x \, \sqrt{-g} \, 
		\frac{1}{\kappa^2} 
		\biggl[
		\mathring{R}
		- \frac{1}{4\kappa^2}
		\biggr] \, .
\label{EH action}
\end{equation}
Note that this action has a rather peculiar property where the cosmological constant
and Newton constant are seemingly not independent, and both depend on the only
dimensionful scale~$\kappa$ that we introduced by hand. 
However, this degeneracy is broken by multiplying the 
entire action by a dimensionless factor.
This factor is fixed by the coupling natter to gravity
if it does not couple in a Weyl-invariant manner.
In a sense~(\ref{Palatini R2}) can be seen as a gauged version of 
the Einstein-Hilbert theory, where the additional local symmetries are projective one~(\ref{projective transformation}), 
and the Weyl one~(\ref{Weyl trans}).

\section{ADM decomposition}
\label{sec: ADM decomposition}

Hamiltonian analyses of gravity theories are greatly facilitated by
clever choice of variables. For pure gravity theories this need was addressed
by the Arnowitt-Deser-Misner (ADM) decomposition~\cite{Arnowitt:1962hi}.
Here we extend the ADM decomposition to 3-tensor fields, and introduce 
appropriate variables for 3-tensors and their time derivatives that
preserve the general ADM structure of gravitational actions. The
particular~$R^2$ theory whose ADM decomposition we seek is first
conveniently written in terms of the metric and the distortion tensor,
\begin{equation}
S\bigl[ g_{\mu\nu}, \Xi_{\alpha\mu\nu} \bigr]
	=
	\int\! d^4x \, \sqrt{-g} \, \biggl[
		\mathring{R} + 
	\mathring{\nabla}_\mu \Bigl( {\Xi^{\mu \nu}}_{\nu} - {\Xi_{\nu}}^{\nu \mu} \Bigr)
	+ {\Xi^{\mu}}_{\mu \alpha} {\Xi^{\alpha \nu} }_{\nu}
	- {\Xi^{\alpha}}_{\mu \nu} {\Xi^{\mu \nu} }_{\alpha}
		\biggr]^2 \, ,
\end{equation}
or even more conveniently in terms of the metric, the non-metricity, 
and the~$B$-tensor according to~(\ref{R decomp QB}),
\begin{align}
S \bigl[ g_{\mu\nu}, B_{\alpha\mu\nu} , Q_{\alpha\mu\nu} \bigr]
	={}& 
	\int\! d^4x \, \sqrt{-g} \, 
	\biggl[ \mathring{R}  
	- \mathring{\nabla}_\mu {B_{\nu}}^{\nu \mu} 
	- \frac{1}{4} {B^{\mu}}_{\mu \alpha} { B_{\nu}}^{\nu \alpha } 
	+ \frac{1}{4} B_{\mu \nu \alpha } B^{\nu \mu \alpha}
\nonumber \\
&	\hspace{2.5cm}
	+ \frac{1}{4} {Q^{\mu}}_{\mu \alpha} { Q_{\nu}}^{\nu \alpha } 
	- \frac{1}{4} Q_{\mu\nu\alpha} Q^{\nu\mu\alpha} \biggr]^2 \, .
\label{S expanded QB}
\end{align}
The standard ADM decomposition of the metric components is given by,
\begin{equation}
g^{00} = - \frac{1}{N^2} \, ,
\qquad \qquad
g_{0i} = N_i \, ,
\qquad \qquad
g_{ij} = h_{ij} \, ,
\end{equation}
where~$N$ and~$N_i$ are the lapse and shift,
and induces the decomposition of inverse components,
\begin{equation}
g_{00} = - N^2 + N^i N_i \, ,
\qquad \qquad
g^{0i} = \frac{N^i}{N^2} \, ,
\qquad \qquad
g^{ij} = h^{ij} - \frac{N^i N^j}{N^2} \, .
\end{equation}
Here~$h_{ij}$ is a spatial metric, and~$h^{ij}$ its inverse,~$h^{ij}h_{jk} \!=\! \delta^i_k$,
and henceforth spatial indices of ADM variables are 
lowered and raised by~$h_{ij}$ and its inverse. 
Note that the lapse cannot be allowed to vanish,~$N \!\neq\! 0$.
The metric determinant is decomposed as~$\sqrt{-g} \!=\! N \sqrt{h}$, and the 
induced ADM decomposition of the 
Christoffel symbols is given in Table~\ref{GammaADM}, where~$\nabla_i$ 
(without the circumflex and with a spatial index) is a covariant derivative with 
respect to the spatial metric~$h_{ij}$. 
%
\begin{table}[h!]
\setlength{\tabcolsep}{10pt}
\def\arraystretch{2.1}
\centering
\begin{tabular}{ l  r  }
\hline 
\hline
	$\mathring{\Gamma}^0_{00}$
	&	
	$\displaystyle \frac{1}{N} \Bigl( \partial_0 N + N^i \nabla_i N - N^i N^j K_{ij} \Bigr)$
\\
\hline
	$\mathring{\Gamma}^i_{00}$
	&	
	$\displaystyle
		- \frac{N^i}{N} \partial_0 N + h^{ij} \partial_0 N_j
		+ \frac{N^i N^j N^k}{N} K_{jk}
		+ N \nabla^i N - \frac{N^i N^j}{N} \nabla_j N - N^j \nabla^i N_j$
\\
\hline 
	$\mathring{\Gamma}^0_{0i}$
	&	
	$\displaystyle \frac{1}{N} \Bigl( - N^j K_{ij} + \nabla_i N \Bigr)$
\\
\hline
	$\mathring{\Gamma}^0_{ij}$
	&	
	$\displaystyle - \frac{1}{N} K_{ij}$
\\
\hline
	$\mathring{\Gamma}^i_{0j}$
	&	
	$\displaystyle - N {K^i}_j + \frac{N^i N^k}{N} K_{jk} - \frac{N^i}{N} \nabla_j N + \nabla_j N^i$
\\
\hline
	$\mathring{\Gamma}^k_{ij}$
	&	
	$\displaystyle \frac{N^k}{N} K_{ij} + \gamma^k_{ij}$
\\
\hline
\hline
\end{tabular}
\caption{ADM decomposition of Christoffel symbols. The Christoffel symbols on spatial hypersurfaces are given by~$\gamma^{k}_{ij} \!=\! \frac{1}{2} h^{kl} \bigl( \partial_i h_{jl} \!+\! \partial_j h_{il} \!-\! \partial_l h_{ij} \bigr)$ .}
\label{GammaADM}
\end{table}
%

Besides the metric itself, it is necessary to define 
convenient variables for time derivatives of the metric as well.
The extrinsic curvature,
\begin{equation}
K_{ij} = - \frac{1}{2N} \Bigl( \partial_0 h_{ij} - 2 \nabla_{(i} N_{j)} \Bigr) \, .
\label{extr curv def}
\end{equation}
is the appropriate variable for the first time derivative of the spatial metric. 
It is also necessary to introduce an appropriate variable for the second time derivative 
of the metric, that resides within the metric Ricci scalar. We find the convenient 
variable to be,
\begin{equation}
F_{ij} = - \frac{1}{N} \partial_0 K_{ij} - K_{ik} {K^k}_j  + \frac{1}{N} N^k \nabla_k K_{ij}
			+ \frac{2}{N}K_{k(i} \nabla_{j)} N^k - \frac{1}{N}\nabla_i \nabla_j N \, ,
\end{equation}
which is just a shifted form of the variable introduced in~\cite{Buchbinder:1987vp}.
Thus the metric Ricci scalar is decomposed as,
\begin{equation}
\mathring{R} = 2 F + K^2 - K_{ij} K^{ij} + \mathcal{R} \, ,
\label{metricR ADM}
\end{equation}
where~$\mathcal{R}$ is the Ricci scalar with respect to spatial metric~$h_{ij}$.

The ADM decompositions of the rank-3 distortion tensor and its time derivative are given
in Tables~\ref{ADMtable1} and~\ref{ADMtable2}.
%
\begin{table}[h!]
\vskip+0.2cm
\setlength{\tabcolsep}{10pt}
\def\arraystretch{1.2}
\centering
\begin{tabular}{ l  r  }
\hline 
\hline
	$\Xi_{000} \quad$
	&	
	$\begin{matrix}
		\hspace{-2.5cm}
		- N^3 \xi_{000}
		+ N^2 N^i \bigl( \xi_{00i} + \xi_{0i0} + \xi_{i00} \bigr) 
		\\
		\hspace{2cm}
		- N N^i N^j \bigl( \xi_{0ij} + \xi_{i0j} + \xi_{ij0} \bigr)
		+ N^i N^j N^k \xi_{ijk}
		\end{matrix}$
\\
\hline
	$\Xi_{00i}$
	&	
	$N^2 \xi_{00i}
		- N N^j \bigl( \xi_{j0i} + \xi_{0ji} \bigr)
		+ N^j N^k \xi_{jki}$
\\
\hline
	$\Xi_{0i0}$
	&	
	$N^2 \xi_{0i0}
		- N N^j \bigl( \xi_{ji0} + \xi_{0ij} \bigr)
		+ N^j N^k \xi_{jik}$
\\
\hline 
	$\Xi_{i00}$
	&	
	$N^2 \xi_{i00}
		- N N^j \bigl( \xi_{ij0} + \xi_{i0j} )
		+ N^j N^k \xi_{ijk}$
\\
\hline
	$\Xi_{0ij}$
	&	
	$- N \xi_{0ij} 	
		+ N^k \xi_{kij}$
\\
\hline
	$\Xi_{i0j}$
	&	
	$- N \xi_{i0j} 	
		+ N^k \xi_{ikj}$
\\
\hline 
	$\Xi_{ij0}$
	&	
	$- N \xi_{ij0} 	
		+ N^k \xi_{ijk}$
\\
\hline
	$\Xi_{ijk}$
	&	
	$\xi_{ijk}$
\\
\hline
\hline
\end{tabular}
\vskip-2mm
\caption{ADM decomposition of distortion tensor~$\Xi$. The~$\xi$'s are tensors on spatial
hypersurfaces and subscripts~$0$ on them denote their names, while spatial indices
determine their tensorial character.}
\label{ADMtable1}
\end{table}
%
\begin{table}[h!]
\vskip+0.3cm
\setlength{\tabcolsep}{10pt}
\def\arraystretch{1.2}
\centering
\begin{tabular}{ l  r  }
\hline 
\hline
	$\partial_0 \xi_{000}\quad$
	&	
	$N W_{000}
		+ N^i \nabla_i \xi_{000}
		- \bigl( \xi_{00i} + \xi_{0i0} + \xi_{i00} \bigr) \nabla^i N$
\\
\hline
	$\partial_0 \xi_{00i}$
	&	
	$N W_{00i} 
		+ N^j \nabla_j \xi_{00i}
		+ \xi_{00j} \nabla_i N^j
		- \bigl( \xi_{j0i} + \xi_{0ji} \bigr) \nabla^j N
		- \xi_{000} \nabla_i N$
\\
\hline
	$\partial_0 \xi_{0i0}$
	&	
	$N W_{0i0} 
		+ N^j \nabla_j M_{0i0}
		+ \xi_{0j0} \nabla_i N^j
		- \bigl( \xi_{ji0} + \xi_{0ij} \bigr) \nabla^j N
		- \xi_{000} \nabla_i N$
\\
\hline 
	$\partial_0 \xi_{i00}$
	&	
	$N W_{i00} 
		+ N^j \nabla_j \xi_{i00}
		+ \xi_{j00} \nabla_i N^j
		- \bigl( \xi_{ij0} + \xi_{i0j} \bigr) \nabla^j N
		- \xi_{000} \nabla_i N$
\\
\hline
	$\partial_0 \xi_{0ij}$
	&	
	$\begin{matrix}
		N W_{0ij}
		+ N^k \nabla_k \xi_{0ij}
		+ \xi_{0ik} \nabla_j N^k
		+ \xi_{0kj} \nabla_i N^k
		\hspace{3cm}
		\\
		\hspace{5cm}
		- \xi_{0i0} \nabla_j N
		- \xi_{00j} \nabla_i N
		- \xi_{kij} \nabla^k N \end{matrix}$
\\
\hline
	$\partial_0 \xi_{i0j}$
	&	
	$\begin{matrix}
		N W_{i0j}
		+ N^k \nabla_k \xi_{i0j}
		+ \xi_{i0k} \nabla_j N^k
		+ \xi_{k0j} \nabla_i N^k
		\hspace{3cm}
		\\
		\hspace{5cm}
		- \xi_{i00} \nabla_j N
		- \xi_{00j} \nabla_i N
		- \xi_{ikj} \nabla^k N\end{matrix}$
\\
\hline 
	$\partial_0 \xi_{ij0}$
	&	
	$\begin{matrix}
		N W_{ij0}
		+ N^k \nabla_k \xi_{ij0}
		+ \xi_{ik0} \nabla_j N^k
		+ \xi_{kj0} \nabla_i N^k
		\hspace{3cm}
		\\
		\hspace{5cm}
		- \xi_{i00} \nabla_j N
		- \xi_{0j0} \nabla_i N
		- \xi_{ijk} \nabla^k N\end{matrix}$
\\
\hline
	$\partial_0 \xi_{ijk}$
	&	
	$\begin{matrix}
		N W_{ijk}
		+ N^l \nabla_l \xi_{ijk}
		+ \xi_{ijl} \nabla_k N^l
		+ \xi_{ilk} \nabla_j N^l
		+ \xi_{ljk} \nabla_i N^l
		\hspace{1.5cm}
		\\
		\hspace{5cm}
		- \xi_{ij0} \nabla_k N
		- \xi_{i0k} \nabla_j N
		- \xi_{0jk} \nabla_i N
		\end{matrix}$
\\
\hline
\hline
\end{tabular}
\vskip-2mm
\caption{ADM variables for time derivatives of distortion ADM variables.}
\label{ADMtable2}
\end{table}
While these decompositions have a geometric interpretation~\cite{Wald:1984rg}, 
we have derived them
in a purely algebraic manner, requiring that the ADM variables absorb all the factors of 
lapse and shift, apart from the overall one coming from the metric determinant. 
Demanding this was enough to fix the variables, up to field redefinitions 
independent of lapse and shift.
Use of the computer algebra program {\it Cadabra} for symbolic manipulation of tensor 
expressions~\cite{Peeters:2007wn,Peeters:2006kp,Peeters:2018dyg} proved 
invaluable for this task.
We have tested that these variables perform the same task of ADM
decomposition for all the quadratic invariants 
that can be formed by~(\ref{Riemann tensor}) and~(\ref{Ricci tensors}).

While Tables~\ref{ADMtable1} and~\ref{ADMtable2} give the ADM decomposition for 
a general 3-tensor without any index exchange symmetries, for our purposes
it is more convenient
to use the non-metricity tensor and the~$B$-tensor that do have index exchange symmetries.
Their ADM decomposition is given in tables~\ref{ADMnonmetricityTable}
and~\ref{ADMBtensorTable}, and decomposition of their time derivatives
is given in tables~\ref{ADM non-metricity ders}
and~\ref{ADM Btensor ders}.
%
\begin{table}[h!]
\vskip+3mm
\setlength{\tabcolsep}{10pt}
\def\arraystretch{1.2}
\centering
\begin{tabular}{ l  r  }
\hline 
\hline
	$Q_{000} \quad$
	&	
	$- N^3 q_{000}
		+ N^2 N^i \bigl( 2 q_{00i} \!+\! q_{i00} \bigr) 
		- N N^i N^j \bigl( q_{0ij} \!+\! 2q_{i0j} \bigr)
		+ N^i N^j N^k q_{ijk}
		$
\\
\hline
	$Q_{00i}$
	&	
	$N^2 q_{00i}
		- N N^j \bigl( q_{j0i} + q_{0ji} \bigr)
		+ N^j N^k q_{jki}$
\\
\hline 
	$Q_{i00}$
	&	
	$N^2 q_{i00}
		- 2 N N^j q_{i0j}
		+ N^j N^k q_{ijk}$
\\
\hline
	$Q_{0ij}$
	&	
	$- N q_{0ij} 	
		+ N^k q_{kij}$
\\
\hline
	$Q_{i0j}$
	&	
	$- N q_{i0j} 	
		+ N^k q_{ikj}$
\\
\hline
	$Q_{ijk}$
	&	
	$q_{ijk}$
\\
\hline
\hline
\end{tabular}
\vskip-0.2cm
\caption{ADM decomposition of non-metricity tensor~$Q$.}
\label{ADMnonmetricityTable}
\end{table}
%
%
\begin{table}[h!]
\vskip+0.3cm
\setlength{\tabcolsep}{10pt}
\def\arraystretch{1.2}
\centering
\begin{tabular}{ l  r  }
\hline 
\hline
	$B_{00i} \quad$
	&	
	$N^2 b_{00i}
		- N N^j \bigl( b_{j0i} \!+\! b_{0ji} \bigr)
		+ N^j N^k b_{jki}$
\\
\hline
	$B_{0ij}$
	&	
	$- N b_{0ij} 	
		+ N^k b_{kij}$
\\
\hline
	$B_{i0j}$
	&	
	$- N b_{i0j} 	
		+ N^k b_{ikj}$
\\
\hline
	$B_{ijk}$
	&	
	$b_{ijk}$
\\
\hline
\hline
\end{tabular}
\vskip-0.2cm
\caption{ADM decomposition of the projective-invariant $B$-tensor.}
\label{ADMBtensorTable}
\end{table}
%
%
\begin{table}[h!]
\vskip+0.3cm
\setlength{\tabcolsep}{10pt}
\def\arraystretch{1.2}
\centering
\begin{tabular}{ l  r  }
\hline 
\hline
	$\partial_0 q_{000} \quad$
	&	
	$N V_{000} 
		+ N^i \nabla_i q_{000}
		- \bigl( 2 q_{00i} + q_{i00} \bigr) \nabla^{i} N $
\\
\hline
	$\partial_0 q_{00i}$
	&	
	$N V_{00i} + N^j \nabla_j q_{00i}
		+ q_{00j} \nabla_i N^j
		- \bigl( q_{j0i} + q_{0ji} \bigr) \nabla^j N
		- q_{000} \nabla^i N$
\\
\hline 
	$\partial_0 q_{i00}$
	&	
	$N V_{i00} + N^j \nabla_j q_{i00}
		+ q_{j00} \nabla_i N^j
		- 2 q_{i0j} \nabla^j N
		- q_{000} \nabla^i N$
\\
\hline
	$\partial_0 q_{0ij}$
	&	
	$\begin{matrix}
		N V_{0ij} + N^k \nabla_k q_{0ij} 
		+ q_{0ik} \nabla_j N^k + q_{0kj} \nabla_i N^k 
		\\
		\hspace{4.5cm}
		- q_{00i} \nabla_j N - q_{00j} \nabla_i N
		- q_{kij} \nabla^k N
		\end{matrix}$
\\
\hline
	$\partial_0 q_{i0j}$
	&	
	$\begin{matrix}
		N V_{i0j} + N^k \nabla_k q_{i0j} 
		+ q_{i0k} \nabla_j N^k + q_{k0j} \nabla_i N^k 
		\\
		\hspace{4.5cm}
		- q_{i00} \nabla_j N - q_{00j} \nabla_i N
		- q_{ikj} \nabla^k N
		\end{matrix}$
\\
\hline
	$\partial_0 q_{ijk}$
	&	
	$\begin{matrix}
		N V_{ijk} + N^l \nabla_l q_{ijk} + q_{ijl} \nabla_k N^l
		+ q_{ilk} \nabla_j N^l + q_{ljk} \nabla_i N^l 
		\\
		\hspace{4.5cm}
		- q_{i0j} \nabla_k N - q_{i0k} \nabla_j N - q_{0jk} \nabla_i N
		\end{matrix}$
\\
\hline
\hline
\end{tabular}
\vskip-0.2cm
\caption{ADM variables for time derivatives of non-metricity}
\label{ADM non-metricity ders}
\end{table}
%
\begin{table}[h!]
\vskip+0.3cm
\setlength{\tabcolsep}{10pt}
\def\arraystretch{1.2}
\centering
\begin{tabular}{ l  r  }
\hline 
\hline
	$\partial_0 b_{00i}$
	&	
	$N U_{00i} + N^j \nabla_j b_{00i}
		+ b_{00j} \nabla_i N^j
		- \bigl( b_{j0i} + b_{0ji} \bigr) \nabla^j N $
\\
\hline
	$\partial_0 b_{0ij}$
	&	
	$N U_{0ij} + N^k \nabla_k b_{0ij}
		+ b_{0ik} \nabla_j N^k + b_{0kj} \nabla_i N^k
		+ b_{00i} \nabla_j N \hspace{1cm} $
\\
&
	$- b_{00j} \nabla_i N - b_{kij} \nabla^k N$
\\
\hline
	$\partial_0 b_{i0j}$
	&	
	$N U_{i0j} + N^k \nabla_k b_{i0j}
		+ b_{i0k} \nabla_j N^k + b_{k0j} \nabla_i N^k
		- b_{00j} \nabla_i N - b_{ikj} \nabla^k N $
\\
\hline
	$\partial_0 b_{ijk}$
	&	
	$\begin{matrix}
		N U_{ijk} + N^l \nabla_l b_{ijk}
		+ b_{ijl} \nabla_k N^l + b_{ilk} \nabla_j N^l + b_{ljk} \nabla_i N^l
		\\
		\hspace{5cm}
		+ b_{i0j} \nabla_k N - b_{i0k} \nabla_j N - b_{0jk} \nabla_i N
		\end{matrix}$
\\
\hline
\hline
\end{tabular}
\vskip-0.2cm
\caption{ADM variables for time derivatives of $B$-tensor}
\label{ADM Btensor ders}
\end{table}
%
Thus the ADM decompositions of the remaining two parts of the 
action~(\ref{S expanded QB}) are, 
\begin{align}
\MoveEqLeft[3]
	\frac{1}{4} {Q^{\mu}}_{\mu \alpha} { Q_{\nu}}^{\nu \alpha } 
	- \frac{1}{4} Q_{\mu\nu\alpha} Q^{\nu\mu\alpha}
	=
	\frac{1}{2} q_{i 0 j} \bigl( {q_0}^{i j} \!+\! h^{i j} q_{0 0 0}\bigr)
	+ \frac{1}{4} \bigl( h^{i l} h^{j k} \!-\! h^{i j} h^{k l} \bigr) q_{i 0 j} q_{k 0 l}
\nonumber \\
&
	- \frac{1}{2} q_{00i} \bigl( {q^i}_{0 0} \!+\! {q_j}^{j i} \bigr)
	+ \frac{1}{4} \bigl( h^{m i} h^{n k} \!-\! h^{m k} h^{i n} \bigr) h^{j l} q_{m i j} q_{n k l}
	\, .
\label{Q part decomposition}
\\
\MoveEqLeft[3]
- \mathring{\nabla}_\mu {B_{\nu}}^{\nu \mu} 
	- \frac{1}{4} {B^{\mu}}_{\mu \alpha} { B_{\nu}}^{\nu \alpha } 
	+ \frac{1}{4} B_{\mu \nu \alpha } B^{\nu \mu \alpha}
	=
	{U^i}_{0i} 
	+ \nabla^i \bigl( b_{00i} \!-\! {b^j}_{ji} \bigr)
	+ b_{i0j} \bigl( 2 K^{ij} \!-\! h^{ij} K \bigr)
\nonumber \\
&
	- \frac{1}{2} {b_{0}}^{i j} b_{i 0 j}
	- \frac{1}{4} {b^i}_{0 j} {b^j}_{0 i}
	+ \frac{1}{4} {b^i}_{0 i} {b^j}_{0 j}
	+ \frac{1}{2} b_{0 0 i} {b_j}^{j i}
	+ \frac{1}{4} b^{i j k} b_{j i k}
	- \frac{1}{4} {b^i}_{i k} {b_j}^{j k} \, .
\label{B part decomposition}
\end{align}
The two ADM decompositions in~(\ref{Q part decomposition})
and~(\ref{B part decomposition}), together with the decomposition 
in~(\ref{metricR ADM}) are the starting point for the Hamiltonian constraint 
analysis that the following section is devoted to.

\section{Canonical action}
\label{sec: Canonical action}

The ADM decomposition of the metric-affine~$R^2$ action
according to~(\ref{metricR ADM}), (\ref{Q part decomposition})
and~(\ref{B part decomposition}),
\begin{align}
\MoveEqLeft[3]
S \bigl[ N, N_i, h_{ij}, b_{\mu\nu\rho}, q_{\mu\nu\rho} \bigr]
	=
	\int\! d^{4\!}x \, N \sqrt{h} \, \biggl[
	2 F + K^2 - K_{ij} K^{ij} + \mathcal{R} 
	+ {U^i}_{0i} 
\nonumber \\
&
	 - \nabla^i \bigl( {b^j}_{ji} \!-\! b_{00i} \bigr)
	- b_{i0j} \bigl( h^{ij} K \!-\! 2 K^{ij} \bigr)
	- \frac{1}{2} {b_{0}}^{i j} b_{i 0 j}
	- \frac{1}{4} {b^i}_{0 j} {b^j}_{0 i}
	+ \frac{1}{4} {b^i}_{0 i} {b^j}_{0 j}
\nonumber \\
&
	+ \frac{1}{2} b_{0 0 i} {b_j}^{j i}
	+ \frac{1}{4} b^{i j k} b_{j i k}
	- \frac{1}{4} {b^i}_{i k} {b_j}^{j k}
	+ \frac{1}{2} q_{i 0 j} \bigl( {q_0}^{i j} \!+\! h^{i j} q_{0 0 0}\bigr)
	+ \frac{1}{4} {q^i}_{0j} {q^j}_{0i}
\nonumber \\
&	
	- \frac{1}{4} {q^i}_{0i} {q^j}_{0j}
	- \frac{1}{2} q_{00i} \bigl( {q^i}_{0 0} \!+\! {q_j}^{j i} \bigr)
	+ \frac{1}{4} {q^i}_{ik} {q_j}^{jk}
	- \frac{1}{4} q^{ijk} q_{jik}
		\biggr]^2 \, .
\label{ADM decomposed action}
\end{align}
is the starting point for the canonical (Hamiltonian) formulation, and the subsequent 
Dirac constraint analysis that this section is devoted to.

\subsection{Canonical action}
\label{subsec: Canonical action}

We derive the canonical action in several steps starting 
from~(\ref{ADM decomposed action}). First we recognize that there are variable 
transformations for non-metricity ADM 
variables,~$q_{\mu\nu\rho} \! \longrightarrow \! z_{\mu\nu\rho}$, given by,
\begin{align}
&
q_{000} = z_{000} \, ,
\qquad
q_{00i} = z_{00i} \, ,
\qquad
q_{i00} = z_{i00} \, ,
\qquad
q_{i0j} = z_{i0j} \, ,
\nonumber \\
&
q_{0ij} = z_{0ij} - h_{ij} z_{000} \, ,
\qquad
q_{ijk} = z_{ijk} - h_{jk} z_{i00} \, ,
\end{align}
which simplify the action somewhat,
\begin{align}
\MoveEqLeft[3]
S \bigl[ N, N_i, h_{ij}, b_{\mu\nu\rho}, z_{\mu\nu\rho} \bigr]
	=
	\int\! d^{4\!}x \, N \sqrt{h} \, \biggl[
	2 F + K^2 - K_{ij} K^{ij} + \mathcal{R} 
	+ {U^i}_{0i} 
\nonumber \\
&
	- \nabla^i \bigl( {b^j}_{ji} \!-\! b_{00i} \bigr)
	- b_{i0j} \bigl( h^{ij} K \!-\! 2 K^{ij} \bigr)
	- \frac{1}{2} {b_{0}}^{i j} b_{i 0 j}
	- \frac{1}{4} {b^i}_{0 j} {b^j}_{0 i}
	+ \frac{1}{4} {b^i}_{0 i} {b^j}_{0 j}
\nonumber \\
&
	+ \frac{1}{2} b_{0 0 i} {b_j}^{j i}
	+ \frac{1}{4} b^{i j k} b_{j i k}
	- \frac{1}{4} {b^i}_{i k} {b_j}^{j k}
	+ \frac{1}{2} {z_0}^{i j}  z_{i 0 j} 
	+ \frac{1}{4} {z^i}_{0j} {z^j}_{0i}
	- \frac{1}{4} {z^i}_{0i} {z^j}_{0j}
\nonumber \\
&	
	- \frac{1}{2} z_{00i} {z_j}^{j i} 
	+ \frac{1}{4} {z^i}_{ik} {z_j}^{jk}
	- \frac{1}{4} z^{ijk} z_{jik}  
		\biggr]^2 \, .
\end{align}
This variable transformation trivializes the projective symmetry, as the
four fields~$z_{\mu00}$ do not appear in the action. This is why we get no first-class constraints 
associated with projective transformations in the Hamiltonian constraint analysis.

We proceed by defining the extended action~$\mathcal{S}$, where time derivatives are promoted
to independent velocity fields,
\begin{equation}
K_{ij} \longrightarrow \mathcal{K}_{ij} \, ,
\qquad \quad
F_{ij} \longrightarrow \mathcal{F}_{ij} \, ,
\qquad \quad
U_{i0i} \longrightarrow \mathcal{U}_{i0i} \, ,
\end{equation}
and accompanying Lagrange multipliers~$\pi^{ij}, \rho^{ij}, P^{i0j}$ are introduced
to ensure on-shell equivalence,
\begingroup
\allowdisplaybreaks
\begin{align}
\MoveEqLeft[2]
\mathcal{S} \bigl[ N, N_i, h_{ij}, b_{\mu\nu\rho}, z_{\mu\nu\rho},
	\mathcal{K}_{ij}, \mathcal{F}_{ij}, \mathcal{U}_{i0j}, \pi^{ij}, \rho^{ij}, P^{i0j} \bigr]
\nonumber \\
	={}&
	\int\! d^{4\!}x \, \Biggl\{
	N \sqrt{h} \, \biggl[
	2 \mathcal{F} + \mathcal{K}^2 - \mathcal{K}_{ij} \mathcal{K}^{ij} + \mathcal{R} 
	+ {\mathcal{U}^i}_{0i} 
	- \nabla^i \bigl( {b^j}_{ji} \!-\! b_{00i} \bigr)
	- b_{i0j} \bigl( h^{ij} \mathcal{K} \!-\! 2 \mathcal{K}^{ij} \bigr)
\nonumber \\
&
	- \frac{1}{2} {b_{0}}^{i j} b_{i 0 j}
	- \frac{1}{4} {b^i}_{0 j} {b^j}_{0 i}
	+ \frac{1}{4} {b^i}_{0 i} {b^j}_{0 j}
	+ \frac{1}{2} b_{0 0 i} {b_j}^{j i}
	+ \frac{1}{4} b^{i j k} b_{j i k}
	- \frac{1}{4} {b^i}_{i k} {b_j}^{j k}
\nonumber \\
&
	+ \frac{1}{2} {z_0}^{i j}  z_{i 0 j} 
	+ \frac{1}{4} {z^i}_{0j} {z^j}_{0i}
	- \frac{1}{4} {z^i}_{0i} {z^j}_{0j}
	- \frac{1}{2} z_{00i} {z_j}^{j i} 
	+ \frac{1}{4} {z^i}_{ik} {z_j}^{jk}
	- \frac{1}{4} z^{ijk} z_{jik}  
		\biggr]^2 
\nonumber \\
&	
	+ \pi^{ij} \Bigl( \partial_0 h_{ij} 
		+ 2 N \mathcal{K}_{ij} - 2\nabla_{(i} N_{j)} \Bigr)
	+ \rho^{ij} \biggl[ \partial_0 \mathcal{K}_{ij} 
		+ N \mathcal{F}_{ij} 
		+ N \mathcal{K}_{ik} {\mathcal{K}^k}_j 
		- N^k \nabla_k \mathcal{K}_{ij}
\nonumber \\
&
		- 2 \mathcal{K}_{k(i} \nabla_{j)} N^k 
		+ \nabla_i \nabla_j N \biggr]
	+ P^{i0j} \biggl[ 
	\partial_0 b_{i0j}
	- N \mathcal{U}_{i0j} - N^k \nabla_k b_{i0j}
		- b_{i0k} \nabla_j N^k 
\nonumber \\
&
		- b_{k0j} \nabla_i N^k
		+ b_{00j} \nabla_i N + b_{ikj} \nabla^k N 
	\biggr]
\Biggr\} \, .
\end{align}
\endgroup
Our next step is to consider the equations 
descending from variations
with respect to velocity fields,
\begin{equation}
\frac{\delta \mathcal{S}}{ \delta \mathcal{K}_{ij} } \approx 0 \, ,
\qquad \quad
\frac{\delta \mathcal{S}}{ \delta \mathcal{F}_{ij} } \approx 0\, ,
\qquad \quad
\frac{\delta \mathcal{S}}{ \delta \mathcal{U}_{i0j} }   \approx 0 \, ,
\end{equation}
and to solve them for as many  velocity fields as possible 
(this step corresponds to the usual Legendre transform).
We cannot solve for all of them on the account of constraints,
but only for a single component
and we choose to solve for~${\mathcal{U}^i}_{0i}$,
\begin{align}
\MoveEqLeft[2]
{\mathcal{U}^i}_{0i} \approx
	\frac{1}{6} \frac{ { P^{i0}}_i }{ \sqrt{h} }
	- \biggl[ 2 \mathcal{F} + \mathcal{K}^2 - \mathcal{K}_{ij} \mathcal{K}^{ij} + \mathcal{R} 
	- \nabla^i \bigl( {b^j}_{ji} \!-\! b_{00i} \bigr)
	- b_{i0j} \bigl( h^{ij} \mathcal{K} \!-\! 2 \mathcal{K}^{ij} \bigr)
\nonumber \\
&
	- \frac{1}{2} {b_{0}}^{i j} b_{i 0 j}
	- \frac{1}{4} {b^i}_{0 j} {b^j}_{0 i}
	+ \frac{1}{4} {b^i}_{0 i} {b^j}_{0 j}
	+ \frac{1}{2} b_{0 0 i} {b_j}^{j i}
	+ \frac{1}{4} b^{i j k} b_{j i k}
	- \frac{1}{4} {b^i}_{i k} {b_j}^{j k}
\nonumber \\
&
	+ \frac{1}{2} {z_0}^{i j}  z_{i 0 j} 
	+ \frac{1}{4} {z^i}_{0j} {z^j}_{0i}
	- \frac{1}{4} {z^i}_{0i} {z^j}_{0j}
	- \frac{1}{2} z_{00i} {z_j}^{j i} 
	+ \frac{1}{4} {z^i}_{ik} {z_j}^{jk}
	- \frac{1}{4} z^{ijk} z_{jik} \biggr] \, .
\end{align}
Plugging this back into the extended action as a strong equality 
produces the canonical action,
\begingroup
\allowdisplaybreaks
\begin{align}
\MoveEqLeft[2]
\mathscr{S} \bigl[ N, N_i, h_{ij}, b_{\mu\nu\rho}, z_{\mu\nu\rho},
	\mathcal{K}_{ij}, \mathcal{F}_{ij}, \widetilde{\mathcal{U}}_{i0j}, \pi^{ij}, \rho^{ij}, P^{i0j} \bigr]
\nonumber \\
	={}&
	\int\! d^{4\!}x \, \Biggl\{
	\pi^{ij} \partial_0 h_{ij} 
	+ \rho^{ij} \partial_0 \mathcal{K}_{ij} 
	+ P^{i0j} \partial_0 b_{i0j}
	- \frac{1}{3} N { P^{l0}}_l  \biggl[
		\frac{1}{12} \frac{ {P^{i0}}_i }{ \sqrt{h} }
		- \mathcal{K}^2 
		+ \mathcal{K}_{ij} \mathcal{K}^{ij} 
\nonumber \\
&
	- \mathcal{R} 
	+ \nabla^i \bigl( {b^j}_{ji} \!-\! b_{00i} \bigr)
	+ b_{i0j} \bigl( h^{ij} \mathcal{K} \!-\! 2 \mathcal{K}^{ij} \bigr)
	+ \frac{1}{4} {b^i}_{0 j} {b^j}_{0 i}
	- \frac{1}{4} {b^i}_{0 i} {b^j}_{0 j}
	- \frac{1}{2} b_{0 0 i} {b_j}^{j i}
\nonumber \\
&
		- \frac{1}{4} b^{i j k} b_{j i k}
		+ \frac{1}{4} {b^i}_{i k} {b_j}^{j k}
		- \frac{1}{2} {z_0}^{i j}  z_{i 0 j} 
		- \frac{1}{4} {z^i}_{0j} {z^j}_{0i}
		+ \frac{1}{4} {z^i}_{0i} {z^j}_{0j}
		+ \frac{1}{2} z_{00i} {z_j}^{j i} 
		- \frac{1}{4} {z^i}_{ik} {z_j}^{jk}
\nonumber \\
&
		+ \frac{1}{4} z^{ijk} z_{jik} 
	\biggr]
	- N \sqrt{h} \biggl[ 
		- 2 \frac{\pi^{ij} }{ \sqrt{h} }  \mathcal{K}_{ij} 
		- \frac{ \rho^{ij} }{ \sqrt{h} } \mathcal{K}_{ik} {\mathcal{K}^k}_j 
		- \nabla_i \nabla_j \frac{ \rho^{ij} }{ \sqrt{h} }
		+  \nabla_i \Bigl( \frac{ P^{i0j} }{ \sqrt{h} } b_{00j} \Bigr) 
\nonumber \\
&	
		+ \nabla^k \Bigl( \frac{ P^{i0j} }{ \sqrt{h} } b_{ikj} \Bigr)
		 \biggr]
	- N_i \sqrt{h} \biggl[ 
		-2 \nabla_j \Bigl( \frac{ \pi^{ij} }{ \sqrt{h} } \Bigr)
		+ \frac{ \rho^{kl} }{ \sqrt{h} } \nabla^i \mathcal{K}_{kl}
		- 2 \nabla_k \Bigl( \frac{ \rho^{kl} }{ \sqrt{h} } {\mathcal{K}_{l}}^i \Bigr)
\nonumber \\
&
		+ \frac{P^{k0l} }{ \sqrt{h} } \nabla^i b_{k0l}
		- \nabla_k  \Bigl( \frac{ P^{l0k} }{ \sqrt{h} } {b_{l0}}^i  \Bigr) 
		- \nabla_k \Bigl( \frac{ P^{k0l} }{ \sqrt{h} } {b^i}_{0l} \Bigr) 
	\biggr]
	+ N \mathcal{F}_{ij} \Bigl[ \rho^{ij} + \frac{2}{3} h^{ij} { P^{l0}}_l \Bigr]
\nonumber \\
&
	- N \widetilde{U}_{i0j} \Bigl[ P^{i0j} - \frac{1}{3} h^{ij} {P^{k0}}_k \Bigr]
	- N b_{0ij} \Bigl[ \frac{ 1 }{6} { P^{k0}}_k \tensor[]{b}{^{[i}_0^{j]}}  \Bigr]
\Biggr\} \, ,
\end{align}
\endgroup
where~$\widetilde{U}_{i0j}$ is the trace-free part of~$U_{i0j}$,~$h^{ij}\widetilde{U}_{i0j} \!=\!0$.
Here we recognize that the remaining velocity fields
now appear as Lagrange multipliers that generate constraints, 
and that we are allowed to shift these multipliers freely. We do so in a way 
that simplifies the action,
\begin{align}
\mathcal{F}_{ij} \longrightarrow{}&
	\mathcal{F}_{ij}
	-
	\mathcal{K}_{ik} {\mathcal{K}^k}_j
	-
	\frac{1}{N} \nabla_i \nabla_j N
	+
	\frac{N_k}{N} \nabla^k \mathcal{K}_{ij}
	+
	\frac{2}{N} {\mathcal{K}^k}_{(i} \nabla_{j)} N_k
	\, ,
\label{F shift}
\\
\widetilde{U}_{ij} \longrightarrow{}&
	\widetilde{U}_{ij}
	+
	b_{00j} \nabla_i N
    +
    b_{ikj} \nabla^k N
    -
    N^k \nabla_k b_{i0j}
    - 
    b_{i0k} \nabla_j N^k
    -
    b_{k0j} \nabla_i N^k
    \, ,
\label{U shift}
\\
b_{0ij} \longrightarrow{}&
	b_{0ij}
	+
	\frac{1}{2} b_{i]0[j}
	\, ,
\label{b shift}
\end{align}
that now reads,
\begin{align}
\MoveEqLeft[2]
\mathscr{S} \bigl[ N, N_i, h_{ij}, b_{\mu\nu\rho}, z_{\mu\nu\rho},
	\mathcal{K}_{ij}, \mathcal{F}_{ij}, \widetilde{\mathcal{U}}_{i0j}, \pi^{ij}, \rho^{ij}, P^{i0j} \bigr]
\nonumber \\
	={}&
	\int\! d^{4\!}x \, \Biggl\{
	\pi^{ij} \partial_0 h_{ij} 
	+ \rho^{ij} \partial_0 \mathcal{K}_{ij} 
	+ P^{i0j} \partial_0 b_{i0j}
	- \frac{1}{3} N P \biggl[
		\frac{1}{12} \frac{ P }{ \sqrt{h} }
		- \mathcal{K}^2 
		+ 3\mathcal{K}_{ij} \mathcal{K}^{ij} 
\nonumber \\
&
	- \mathcal{R} 
	+ \nabla^i \bigl( {b^j}_{ji} \!-\! b_{00i} \bigr)
	+ b_{i0j} \bigl( h^{ij} \mathcal{K} \!-\! 2 \mathcal{K}^{ij} \bigr)
	+ \frac{1}{4} b_{i 0 j} \tensor[]{b}{^{[j}_0^{i]}}
	- \frac{1}{4} b^2
	- \frac{1}{2} b_{0 0 i} {b_j}^{j i}
\nonumber \\
&
		- \frac{1}{4} b^{i j k} b_{j i k}
		+ \frac{1}{4} {b^i}_{i k} {b_j}^{j k}
		- \frac{1}{2} {z_0}^{i j}  z_{i 0 j} 
		- \frac{1}{4} {z^i}_{0j} {z^j}_{0i}
		+ \frac{1}{4} {z^i}_{0i} {z^j}_{0j}
		+ \frac{1}{2} z_{00i} {z_j}^{j i} 
		- \frac{1}{4} {z^i}_{ik} {z_j}^{jk}
\nonumber \\
&
		+ \frac{1}{4} z^{ijk} z_{jik} 
	\biggr]
	- N \sqrt{h} \biggl[ 
		\frac{2}{3} \nabla^k \nabla_k \Bigl( \frac{P}{ \sqrt{h} } \Bigr)
		- 2 \frac{\pi^{ij} }{ \sqrt{h} }  \mathcal{K}_{ij} 
		+ \frac{1}{3} \nabla^i \Bigl( \frac{ P }{ \sqrt{h} } \bigl[ b_{00i} - {b^j}_{ji} \bigr] \Bigr) 
		 \biggr]
\nonumber \\
&	
	- N_i \sqrt{h} \biggl[ 
		-2 \nabla_j \Bigl( \frac{ \pi^{ij} }{ \sqrt{h} } \Bigr)
		+ \frac{1}{3} \frac{P }{ \sqrt{h} } \nabla^i \bigl( b - 2 \mathcal{K} \bigr)
		+ \frac{2}{3} \nabla_j \Bigl( \frac{P}{ \sqrt{h} } \bigl[ 2 \mathcal{K}^{ij} - \tensor[]{b}{^{(i}_0^{j)}} \bigr] \Bigr)
	\biggr]
\nonumber \\
&
	+ N \mathcal{F}_{ij} \Bigl[ \rho^{ij} + \frac{2}{3} h^{ij} P \Bigr]
	- N \widetilde{U}_{i0j} \Bigl[ P^{i0j} - \frac{1}{3} h^{ij} P \Bigr]
	- N b_{0ij} \Bigl[ \frac{ 1 }{6} P \tensor[]{b}{^{[i}_0^{j]}}  \Bigr]
\Biggr\} \, ,
\label{pre-canonical action}
\end{align}
and is further simplified by introducing a shorthand notation for the traces,
\begin{equation}
\pi \equiv {\pi^i}_i \, ,
\qquad \quad
\rho \equiv {\rho^i}_i \, ,
\qquad \quad
P \equiv {P^{i0}}_i \, ,
\qquad \quad
b \equiv {b^i}_{0i} \, .
\end{equation}
Next we note that there are still some auxiliary fields left in the canonical 
action, that appear non-linearly, and without associated canonical momenta.
These are first all the~$z$'s that we can solve for on-shell.
We do so by considering the equations 
of motion produced by varying with respect to these variables. The first pair of equations,
\begin{align}
0 \approx{}&
	\frac{\delta \mathcal{S}' }{ \delta z_{00i} }
	=
	- \frac{1}{6} N P {z_j}^{ji} \, ,
\\
0 \approx{}&
	\frac{\delta \mathcal{S}' }{ \delta z_{ijk} }
	=
	- \frac{1}{6} N P \Bigl[ z^{(jk)i} - {z_l}^{l(k} h^{j)i} + {z_{00}}^{(k} h^{j)i} \Bigr] \, ,
\end{align}
implies~$z_{00i} \!\approx\!0$ and~$z_{ijk} \!\approx\! 0$, while the second set,
\begin{align}
0 \approx{}&\frac{\delta \mathcal{S}}{\delta z_{0ij} }
	= \frac{1}{12} N P \Bigl[ \tensor[]{z}{^i_0^j} + \tensor[]{z}{^j_0^i} \Bigr] \, ,
\\
0 \approx{}&\frac{\delta \mathcal{S}}{\delta z_{i0j} }
	= \frac{1}{6} N P \Bigl[ \tensor[]{z}{^j_0^i} 
		- h^{ij} {z^k}_{0k} + {z_0}^{ij}
	\Bigr] \, ,
\end{align}
implies~$z_{0ij} \!\approx\! 0$ and~$z_{i0j} \!\approx\!0$. 
This means that on-shell all the components of non-metricity vanish, apart from~$z_{\mu00}$ 
which are pure gauge and can be chosen arbitrarily.
It is also that~$b_{00i}$ and~$b_{ijk}$ are auxiliary fields
that we can solve for, 
which we do by considering equations of motion coming from their variations,
\begin{align}
0 \approx{}&
\frac{ \delta \mathcal{S}'' }{ \delta b_{00i} } =
	\frac{N}{6} P {b_j}^{ji}
	- \frac{N \sqrt{h}}{3} \nabla^i \Bigl( \frac{ P }{ \sqrt{h} } \Bigr)
	\, ,
\\
0 \approx{}&
\frac{ \delta \mathcal{S}'' }{ \delta b_{ijk} } =
		\frac{N}{6} P \Bigl( 
	b^{[kj]i} - {b_m}^{m[k} h^{j]i} + {b_{00}}^{[k} h^{j]i} \Bigr)
	+ \frac{N \sqrt{h} }{3}  h^{i[j} \nabla^{k]} \Bigl( \frac{ P }{ \sqrt{h} } \Bigr) \, .
\end{align}
These imply that on-shell,
\begin{equation}
b_{00i} \approx
	- \frac{\sqrt{h} }{P} \nabla_i \Bigl( \frac{ P }{ \sqrt{h} } \Bigr)
	 \, ,
\qquad \qquad
b^{kij} \approx
	2 \frac{\sqrt{h}}{P} h^{k[i} \nabla^{j]} \Bigl( \frac{ P }{ \sqrt{h} } \Bigr)
	\, .
\end{equation}
Plugging these back into the action~(\ref{pre-canonical action}) produces the 
final form of the canonical action,
\begin{align}
\MoveEqLeft[5]
\mathscr{S}\bigl[ ( h_{ij}, \pi^{ij} ), (\mathcal{K}_{ij}, \rho^{ij} ), (b_{i0j}, P^{i0j}) ;
	N, N_i,  \mathcal{F}_{ij}, \widetilde{\mathcal{U}}_{i0j}, b_{0ij} \bigr]
\nonumber \\
	={}& \int\! d^{4\!}x \, \biggl[
		\pi^{ij} \partial_0 h_{ij}
		+ \rho^{ij} \partial_0 \mathcal{K}_{ij}
		+ P^{i0j} \partial_0 b_{i0j}
		- N \mathcal{H} - N_i \mathcal{H}^i
\nonumber \\
&	\hspace{4.5cm}
	- N \mathcal{F}_{ij} \mathcal{C}^{ij}
	- N \widetilde{U}_{i0j} \Psi^{ij}
	- N b_{0ij}\Phi^{ij}
		\biggr] \, .
\label{final canonical action}
\end{align}
In the list of arguments of the canonical action above pairs of canonically conjugate variables are 
indicated by parentheses, while the ones following the semicolon are Lagrange multipliers.
The Poisson brackets between canonical pairs are fixed by the symplectic part of the action,
while the Hamiltonian is composed out of constraints only, that are given by,
\begin{align}
\mathcal{H} ={}&
	\sqrt{h} \biggl[ \frac{1}{3} \frac{ P }{ \sqrt{h} } \biggl(
	\frac{1}{12} \frac{ P }{ \sqrt{h} }
	- \mathcal{K}^2 
	+ 3 \mathcal{K}_{ij} \mathcal{K}^{ij} 
	- \mathcal{R} 
	+ b_{i0j} \bigl( h^{ij} \mathcal{K} \!-\! 2 \mathcal{K}^{ij} \bigr)
	+ \frac{1}{4} b_{i0 j} \tensor[]{b}{^{(i}_0^{j)}}
	- \frac{1}{4} b^2
	\biggr)
\nonumber \\
&	\hspace{1cm}
	- 2 \frac{\pi^{ij} }{ \sqrt{h} }  \mathcal{K}_{ij} 
		+ \frac{2}{3} \nabla_i \nabla^i \Bigl( \frac{ P }{ \sqrt{h} } \Bigr)
		- \frac{1}{2}\frac{\sqrt{h}}{ P }
			\nabla_i \Bigl( \frac{ P }{\sqrt{h}} \Bigr)
			\nabla^i \Bigl( \frac{ P }{\sqrt{h}} \Bigr)
		\biggr] 
		\, ,
\label{Hamiltonian constraint}
\\
\mathcal{H}^{i}
	={}&
	\sqrt{h} \biggl[ 
		-2 \nabla_j \Bigl( \frac{ \pi^{ij} }{ \sqrt{h} } \Bigr)
		- \frac{1}{3} \frac{ P }{\sqrt{h}} \nabla^i \bigl( 2\mathcal{K} - b \bigr)
		+ \frac{2}{3} \nabla_j \Bigl( \frac{ P }{ \sqrt{h} } \bigl[ 2 \mathcal{K}^{ji} - \tensor[]{b}{^{(i}_0^{j)}} \bigr] \Bigr)
	\biggr]	
	\, ,
\label{momentum constraint}
\\
C^{ij} \! ={}&
	- \rho^{ij} - \frac{2}{3} h^{ij} P
	\, ,
\qquad
\Psi^{ij} =
	{H^{ij}}_{kl} P^{k0l} 
	+
	{\overline{H}^{ij}}_{kl} P^{k0l} 
	\, ,
\qquad
\Phi^{ij} =
	\frac{1}{6} P \overline{H}^{ijkl} b_{k0l}
	\, ,
\label{extra constraints}
\end{align}
where for the last two constraints we introduced a shorthand notation
for anti-symmetric and symmetric traceless projectors,
\begin{equation}
{\overline{H}^{ij}}_{kl} = \delta^{[i}_k \delta^{j]}_l \, ,
\qquad \qquad
{H^{ij}}_{kl} = \delta^{(i}_k \delta^{j)}_l - \frac{1}{3} h^{ij} h_{kl} \, .
\end{equation}
The constraint in~(\ref{momentum constraint}) can be seen as the
momentum constraint, while the combination of constraints 
in~(\ref{Hamiltonian constraint}) and~(\ref{extra constraints}),
$(\mathcal{H} \!+\! \mathcal{F}_{ij}
    \!+\! \widetilde{U}_{ij} \Psi^{ij} \!+\! b_{0ij} \Phi^{ij})$,
can be seen to be the Hamiltonian constraint as it collects all the
terms proportional to lapse in the canonical 
action~(\ref{final canonical action}).~\footnote{Some care needs to be
given when defining Hamiltonian and momentum constraints as off-shell
generators of diffeomorphisms, where pieces proportional to the
constraints in~(\ref{extra constraints}) have to be added to them.
However, this makes no difference for the on-shell analysis of constraints
we perform in the following section.
}
The appearance of these constraints is standard in the metric 
formulation of gravity, while the appearance of extra constraints
in~(\ref{extra constraints}) that vanish on-shell on their own
is associated to the disformation sector
descending from the affine connection.

\section{Constraint analysis}
\label{sec: Constraint analysis}

The primary constraints~(\ref{Hamiltonian constraint})--(\ref{extra constraints}) identified in the preceding section are the first
step of the Dirac-Bergman algorithm~\cite{DiracBook} that we employ
in order to determine the number of propagating degrees of freedom in the
theory. The algorithm requires (i) to check for the conservation of constraints and to identify possible secondary and further generations 
of constraints descending from it, and (ii) to compute the Poisson 
brackets between all the constraints to classify them as first- or 
second-class. This section is devoted to performing this analysis.

We find it most convenient to split the constraints, apart from the Hamiltonian
and momentum constraints, into three sectors according to their tensor
character: anti-symmetric, symmetric traceless, and trace sectors.
The brackets of primary constraints between different sectors 
all vanish.
We use the ambiguities in the on-shell definition of constraints to maintain
this property of disjoint constraint sectors through all generations of constraints.
All the constraints we find have weakly vanishing brackets with the momentum constraint~(\ref{momentum constraint}). 
This is not the case with the Hamiltonian constraint~(\ref{Hamiltonian constraint}), 
which is the reason why secondary constraints are generated.

\subsection{Primary constraints}
\label{subsec: Primary constraints}
 
The primary constraints~(\ref{extra constraints}) are split into three sectors 
according to their tensor properties. The primary constraints from any given 
sector commute with constraints of other sectors.

\noindent {\bf Anti-symmetric sector.}
The two anti-symmetric primary constraints are,
\begin{equation}
\Psi^{[ij]} = {\overline{H}^{ij}}_{kl} P^{k0l} \, ,
\qquad \quad
\Phi^{ij} = \frac{1}{24} P \overline{H}^{ijkl} b_{i0j}
\, .
\label{anti-symmetric primary}
\end{equation}
They are mutually second-class,
\begin{align}
\bigl\{ \Psi^{[ij]} (t,\vec{x}) , \Psi^{[kl]}(t,\vec{x}^{\,\prime} ) \bigr\}
	\approx{}& 
	0
	\, ,
\label{anti-sym bracket 1}
\\
\bigl\{ \Psi^{[ij]} (t,\vec{x}) , \Phi^{kl}(t,\vec{x}^{\,\prime} ) \bigr\}
	\approx{}& 
	- \frac{1}{6} P \overline{H}^{ijkl} \delta^3(\vec{x} \!-\! \vec{x}^{\,\prime} )
	\, ,
\label{anti-sym bracket 2}
\\
\bigl\{ \Phi^{ij} (t,\vec{x}) , \Phi^{kl}(t,\vec{x}^{\,\prime} ) \bigr\}
	\approx{}&
	0
	\, .
\label{anti-sym bracket 3}
\end{align}

\medskip

\noindent{\bf Symmetric traceless sector.}
The two primary constraints of the symmetric traceless sector are,
\begin{equation}
\Psi^{(ij)} =
	{H^{ij}}_{kl} P^{k0l}
	\, ,
\qquad \qquad
\mathcal{C}^{ij} =
    { H^{ij} }_{kl} C^{kl}
    =
	- 
	{H^{ij}}_{kl} \rho^{kl}
	\, .
\label{symmetric primary}
\end{equation}
All the brackets between them vanish,
\begin{align}
\bigl\{ \Psi^{(ij)} (t,\vec{x}) , \Psi^{(kl)}(t,\vec{x}^{\,\prime} ) \bigr\}
	\approx{}& 
	0
	\, ,
\\
\bigl\{ \Psi^{(ij)} (t,\vec{x}) , \mathcal{C}^{kl}(t,\vec{x}^{\,\prime} ) \bigr\}
	\approx{}& 
	0
	\, ,
\\
\bigl\{ \mathcal{C}^{ij} (t,\vec{x}) , \mathcal{C}^{kl}(t,\vec{x}^{\,\prime} ) \bigr\}
	\approx{}& 
	0
	\, .
\end{align}

\medskip

\noindent{\bf Trace sector.}
There is a single scalar primary constraint,
\begin{equation}
\mathcal{C} 
    = h_{ij} C^{ij}
    = - \rho - 2 P \, .
\label{trace primary}
\end{equation}
%

\subsection{Conservation of primary constraints}
\label{subsec: Conservation of primary constraints}

\noindent {\bf Anti-symmetric sector.} 
The conservation of primary constraints~(\ref{anti-symmetric primary}) 
does not generate secondary constraints. Rather, the two constraints are mutually 
second-class according to their 
brackets~(\ref{anti-sym bracket 1})--(\ref{anti-sym bracket 3}),
and their conservation determines the Lagrange 
multipliers~$\widetilde{U}_{i]0[j}$ and~$b_{0ij}$
on-shell.

\medskip

\noindent{\bf Symmetric traceless sector.}
The conservation of the two primary constraints in~(\ref{symmetric primary}),
\begin{equation}
0 \approx
	\partial_0 \Psi^{(ij)}
	\approx N \Theta^{ij}
	\, ,
\qquad \quad
0 \approx
	\partial_0 \mathcal{C}^{ij}
	\approx N \mathcal{X}^{ij}
	\, ,
\end{equation}
generates two secondary symmetric traceless constraints,
\begin{equation}
\Theta^{ij}
	=
	- \frac{1}{6} P
	H^{ijkl} b_{k0l}
	\, ,
\qquad \quad
\mathcal{X}^{ij}
	=
	{H^{ij}}_{kl}
	\biggl( - 2 \pi^{kl} + \frac{2}{3} P \mathcal{K}^{kl} \biggr)
	\, .
\label{symmetric secondary}
\end{equation}
The two primary and the two secondary constraints now form a complete set of 
second-class constraints, as the remaining brackets of the sector are,
\begin{align}
\bigl\{ \Theta^{ij} (t,\vec{x}) , \Theta^{kl}(t,\vec{x}^{\,\prime} ) \bigr\}
	\approx{}& 
	0
	\, ,
\label{symm comm first}
\\
\bigl\{ \Theta^{ij} (t,\vec{x}) , \mathcal{X}^{kl}(t,\vec{x}^{\,\prime} ) \bigr\}
	\approx{}& 
	- \frac{1}{9} P b H^{ijkl} \delta^3(\vec{x} \!-\! \vec{x}^{\,\prime})
	\, ,
\\
\bigl\{ \mathcal{X}^{ij} (t,\vec{x}) , \mathcal{X}^{kl}(t,\vec{x}^{\,\prime} ) \bigr\}
	\approx{}& 
	0
	\, ,
\\
\bigl\{ \Theta^{ij} (t,\vec{x}) , \Psi^{(kl)}(t,\vec{x}^{\,\prime} ) \bigr\}
	\approx{}& 
	- \frac{1}{6}
	P H^{ijkl} 
	\delta^3(\vec{x} \!-\! \vec{x}^{\,\prime})
	\, ,
\\
\bigl\{ \Theta^{ij} (t,\vec{x}) , \mathcal{C}^{kl}(t,\vec{x}^{\,\prime} ) \bigr\}
	\approx{}& 
	0
	\, ,
\\
\bigl\{ \mathcal{X}^{ij} (t,\vec{x}) , \Psi^{(kl)}(t,\vec{x}^{\,\prime} ) \bigr\}
	\approx{}& 
	\frac{2}{3} 
	P
	H^{ijkl}
	\delta^3(\vec{x} \!-\! \vec{x}^{\,\prime})
	\, ,
\\
\bigl\{ \mathcal{X}^{ij} (t,\vec{x}) , \mathcal{C}^{kl}(t,\vec{x}^{\,\prime} ) \bigr\}
	\approx{}& 
	-  \frac{ 2}{3} P H^{ijkl}
	\delta^3(\vec{x} \!-\! \vec{x}^{\,\prime})
	\, ,
\label{symm comm last}
\end{align}

\medskip

\noindent{\bf Trace sector.} 
The conservation of the primary constraint~(\ref{trace primary}),
\begin{equation}
0
\approx
\partial_0 \mathcal{C}
	\approx
	N
	\mathcal{X}
	\, .
\label{trace secondary}
\end{equation}
generates a secondary scalar constraint,~\footnote{Note that we could have dropped
the last term from the secondary constraint in~(\ref{trace secondary}) as it is 
proportional to the second primary constraint in~(\ref{symmetric primary}) 
of the symmetric sector, and on-shell there is no difference.
However, in that case we would not respect the requirement
that different sectors commute between themselves, which simplifies the structure
significantly.}
\begin{equation}
\mathcal{X} \equiv 
	-
	2 \pi
	+ 
	\frac{P}{3}  \bigl( 2 \mathcal{K} - b \bigr)
	-
	\mathcal{K}_{ij} {H^{ij}}_{kl} \rho^{kl}
	\, .
\label{scalar X constraint}
\end{equation}
The brackets between the primary and secondary constraints vanishes,
\begin{equation}
\bigl\{ \mathcal{C}(t,\vec{x}) , \mathcal{X}(t,\vec{x}^{\,\prime}) \bigr\} \approx 0 \, .
\end{equation}
%

\subsection{Conservation of secondary constraints}
\label{subsec: Conservation of secondary constraints}

\noindent{\bf Symmetric traceless sector.}
The brackets~(\ref{symm comm first})--(\ref{symm comm first})
imply that the four symmetric sector constraints are
mutually second-class. Therefore, rather than generating tertiary constraints,
their conservation determines the Lagrange multipliers associated
to the primary constraints.

\medskip

\noindent{\bf Trace sector.} 
the conservation of the scalar sector secondary constraint,
\begin{equation}
\partial_0 \mathcal{X} \approx 0 \, ,
\end{equation}
does not generate further constraints. This implies that the two constraints in the
scalar sector are first-class.

\subsection{Summary of the constraint structure}
\label{subsec: Summary of the constraint structure}

In addition to the Poisson brackets worked out in 
sections~\ref{subsec: Primary constraints}--\ref{subsec: Conservation of secondary constraints} it is also true that the brackets between different sectors of constraints vanish
weakly (on-shell), 
and that the momentum constraint has weakly vanishing brackets with 
all the rest.
This leaves us with a set of~$N_{\rm 1st}\!=\!6$ 
first-class constraints
after the Lagrange multipliers are determined
on-shell,~$(\mathcal{H} \!+\! \mathcal{F}_{ij}
    \!+\! \widetilde{U}_{ij} \Psi^{ij} \!+\! b_{0ij} \Phi^{ij})$,~$\mathcal{H}^i$,~$\mathscr{C}$,
    and~$\mathcal{X}$,
and two blocks of second-class constraints in the anti-symmetric 
and symmetric sectors. 
The constraint structure is summarized in
Table~\ref{constraint summary}.
%
\begin{table}[h!]
\setlength{\tabcolsep}{10pt}
\def\arraystretch{1.2}
\centering
\begin{tabular}{ l  c  c  c }
\hline \hline
sector
&
anti-symmetric
&
traceless symmetric
&
scalar
\\
\hline \hline
primary:
&
$\Psi^{[ij]}, \Phi^{ij} $ 
&
$\mathcal{C}^{ij}, \Psi^{(ij)}$
&
$\mathcal{C}$
\\
&
$\Downarrow$
&
$\Downarrow$
&
$\Downarrow$
\\
secondary: 
&
\Large $ \times$
&
$\Theta^{ij}, \mathcal{X}^{ij}$
&
$\mathcal{X}$
\\
&
&
$\Downarrow$
&
$\Downarrow$
\\
tertiary:
&
&
\Large $ \times$
&
\Large $ \times$
\\
\hline \hline
\end{tabular}
\vskip-2mm
\caption{Generations of constraints across three sectors.}
\label{constraint summary}
\end{table}
%
In total there are~$N_{\rm 2nd}\!=\!26$ second-class constraints
and~$N_{\rm 1st}\!=\!6$ first-class constraints.
Given that there are~$N_{\rm can}\!=\!42$ canonical variables
(not including Lagrange multipliers) we can infer the number of physical 
propagating degrees of freedom,
\begin{equation}
N_{\rm phy}
	=
	\frac{1}{2}
	\Bigl(
	N_{\rm can}
	-
	N_{\rm 2nd}
	-
	2 N_{\rm 1st}
	\Bigr)
	=
	2
	\, .
\end{equation}
This conclusion agrees with Sec.~\ref{sec: Equivalence to Einstein-Hilbert}
where the field redefinition method was used to infer that
the number of propagating degrees of freedom is two.

\section{Reducing phase space}
\label{sec: Reducing phase space}

The Dirac constraint analysis of the preceding section allowed us to 
identify all the constraints of the theory, and to classify them 
by character into first- and second-class ones. In this section we 
reduce the phase space of the theory by explicitly solving
the~$N_{\rm 2nd}\!=\!26$ second-class constraints for 26 canonical 
variables. This procedure generally allows better insight into 
the physical content of the theory.
We first solve the anti-symmetric sector constraints in~(\ref{anti-symmetric primary}) for,
\begin{equation}
{\overline{H}^{ij}}_{kl} P^{k0l} \approx 0 \, ,
\qquad \qquad
	\overline{H}^{ijkl} b_{k0l} \approx 0 \, ,
\label{antisymmetric solutions}
\end{equation}
and the symmetric traceless sector constraints in~(\ref{symmetric primary})
and~(\ref{symmetric secondary}) for,
\begin{align}
	{H^{ij}}_{kl} P^{k0l} \approx 0 \, ,
\qquad \
	{H^{ij}}_{kl} \rho^{kl} \approx 0 \, ,
\qquad \
	H^{ijkl} b_{k0l} \approx 0 \, ,
\qquad \
	H^{ijkl} \mathcal{K}_{kl} \approx \frac{3}{P} {H^{ij}}_{kl} \pi^{kl} \, .
\label{symmetric solutions}
\end{align}
The trace sector contains only first-class constraints, which we are not allowed 
to solve for.  Furthermore, in the reduced formulation
it is convenient to use traces as variables,
\begin{align}
&
\mathcal{K} = h^{ij} \mathcal{K}_{ij} \longrightarrow \mathscr{K} \, ,
\qquad \quad
\rho = h_{ij} \rho^{ij} \longrightarrow \mathscr{M} \, ,
\nonumber \\
& b = h^{ij} b_{i0j} \longrightarrow \mathscr{B} \, ,
\qquad \quad
P = h_{ij} P^{i0j} \longrightarrow \mathscr{P} \, .
\label{trace variables}
\end{align}
Given the solutions in~(\ref{antisymmetric solutions})
and~(\ref{symmetric solutions}), and the new variables in~(\ref{trace variables})
the reduced phase space canonical action is given by,
\begingroup
\allowdisplaybreaks
\begin{align}
\MoveEqLeft[0.5]
\mathscr{S}_{\rm red}
	\bigl[ h_{ij}, \pi^{ij} , \mathscr{K}, \mathscr{M} , \mathscr{B} , \mathscr{P} ,
	N, N_i,  \mathscr{F} \bigr]
\nonumber \\
	&= \! \int\! d^{4\!}x \, \Biggl\{
		\biggl[
		\pi^{ij} 
		+ \frac{ \mathscr{M} }{ \mathscr{P} } {H^{ij}}_{kl} \pi^{kl}
		+ \frac{ \mathscr{M} \mathscr{K} h^{ij}  }{9}
		+ \frac{ \mathscr{P} \mathscr{B} h^{ij} }{9}
			\biggr] \partial_0 h_{ij}
		+ \frac{ \mathscr{M} \partial_0 \mathscr{K} }{3}
		+ \frac{ \mathscr{P} \partial_0 \mathscr{B} }{3}
\nonumber \\
&
	- N 
	\sqrt{h} \biggl[ 
	3 \frac{ \sqrt{h} }{ \mathscr{P} } \frac{ \pi^{ij} }{ \sqrt{h} } H_{ijkl} \frac{ \pi^{kl} }{ \sqrt{h} }
	- 
	\frac{ 2\mathscr{K} }{3} \frac{\pi }{ \sqrt{h} }
	-
	\frac{ \mathscr{P} }{ \sqrt{h} } \frac{ \mathcal{R} }{3} 
	+
	\frac{ \mathscr{P} }{ \sqrt{h} } 
	\frac{ \mathscr{B} \bigl( 2\mathscr{K} \!-\! \mathscr{B} \bigr) }{18}
	+
	\frac{1}{36} \Bigl( \frac{ \mathscr{P} }{ \sqrt{h} } \Bigr)^{\!2}
\nonumber \\
&	
	+ 
	\frac{2}{3} \nabla_i \nabla^i \Bigl( \frac{ \mathscr{P} }{ \sqrt{h} } \Bigr)
	- 
	\frac{1}{2}\frac{\sqrt{h}}{ \mathscr{P} }
			\nabla_i \Bigl( \frac{ \mathscr{P} }{\sqrt{h}} \Bigr)
			\nabla^i \Bigl( \frac{ \mathscr{P} }{\sqrt{h}} \Bigr)
		\biggr] 
	- N_i
	\sqrt{h} \biggl[ 
	2 \nabla_j \Bigl( \frac{ \pi^{ij} }{ \sqrt{h} }  \Bigr)
	- 
	\frac{4}{3} \nabla^i \Bigl( \frac{\pi}{ \sqrt{h} } \Bigr)
\nonumber \\
&
	+ 
	\frac{2}{9} 
		\bigl( 2 \mathscr{K} \!-\! \mathscr{B} \bigr)
		\nabla^i \Bigl( \frac{ \mathscr{P} }{ \sqrt{h} } \Bigr)
	- 
	\frac{1}{9} \frac{ \mathscr{P} }{ \sqrt{h} } 
		\nabla^i \bigl( 2 \mathscr{K} \!-\! \mathscr{B} \bigr)
	\biggr]	
	+ \frac{N}{3} \mathscr{F} \bigl( \mathscr{M} \!+\! 2 \mathscr{P} \bigr)
		\Biggr\} \, .
\end{align}
\endgroup
The conservation of the primary constraint here generates the phase space reduced
secondary constraint~(\ref{scalar X constraint}),
\begin{equation}
\partial_0 \bigl( - \mathscr{M} \!-\! 2 \mathscr{P} \bigr)
	\approx
	N
	\biggl[
	- 2 \pi + \frac{\mathscr{P}}{3} \bigl( 2 \mathscr{K} \!-\! \mathscr{B} \bigr)
	\biggr]
	\approx 0 \, .
\end{equation}
We introduce this secondary constraint explicitly
into the action with its own accompanying Lagrange multiplier~$\mathscr{G}$, 
as is generally 
allowed~\cite{Henneaux:1992ig}. This step simplifies the analysis of this section, 
as it allows to absorb some terms into the new multiplier, and 
thus to eliminate both~$\mathscr{K}$ and~$\mathscr{B}$ from the reduced Hamiltonian 
and momentum constraints,
\begin{align}
\MoveEqLeft[0.5]
\mathscr{S}_{\rm red}
	\bigl[ h_{ij}, \pi^{ij} , \mathscr{K}, \mathscr{M} , \mathscr{B} , \mathscr{P} ,
	N, N_i,  \mathscr{F} , \mathscr{G} \bigr]
\nonumber \\
	&= \! \int\! d^{4\!}x \, \Biggl\{
		\biggl[
		\pi^{ij} 
		+ \frac{ \mathscr{M} }{ \mathscr{P} } {H^{ij}}_{kl} \pi^{kl}
		+ \frac{ \mathscr{M} \mathscr{K} h^{ij}  }{9}
		+ \frac{ \mathscr{P} \mathscr{B} h^{ij} }{9}
			\biggr] \partial_0 h_{ij}
		+ \frac{ \mathscr{M} \partial_0 \mathscr{K} }{3}
		+ \frac{ \mathscr{P} \partial_0 \mathscr{B} }{3}
\nonumber \\
&
	- N 
	\sqrt{h} \biggl[ 
	3 \frac{ \sqrt{h} }{ \mathscr{P} } 
		\biggl( \frac{ \pi^{ij} }{ \sqrt{h} } \frac{ \pi_{ij} }{ \sqrt{h} }
			- \frac{ \pi }{ \sqrt{h} } \frac{ \pi }{ \sqrt{h} } \biggr)
	-
	\frac{ \mathscr{P} }{ \sqrt{h} } \frac{ \mathcal{R} }{3} 
	+
	\frac{1}{36} \Bigl( \frac{ \mathscr{P} }{ \sqrt{h} } \Bigr)^{\!2}
	+ 
	\frac{2}{3} \nabla_i \nabla^i \Bigl( \frac{ \mathscr{P} }{ \sqrt{h} } \Bigr)
\nonumber \\
&	
	- 
	\frac{1}{2}\frac{\sqrt{h}}{ \mathscr{P} }
			\nabla_i \Bigl( \frac{ \mathscr{P} }{\sqrt{h}} \Bigr)
			\nabla^i \Bigl( \frac{ \mathscr{P} }{\sqrt{h}} \Bigr)
		\biggr] 
	- N_i
	\sqrt{h} \biggl[ 
	2 \nabla_j \Bigl( \frac{ \pi^{ij} }{ \sqrt{h} }  \Bigr)
	- 
	2 \nabla^i \Bigl( \frac{\pi}{ \sqrt{h} } \Bigr)
\\
&
	+
	2 \frac{\pi}{ \sqrt{h} } 
		\Bigl( \frac{ \mathscr{P} }{ \sqrt{h} } \Bigr)^{\!-1}
		\nabla^i \Bigl( \frac{ \mathscr{P} }{\sqrt{h}}  \Bigr)
	\biggr]	
	+ 
	\frac{N}{3} \mathscr{F} \bigl( \mathscr{M} \!+\! 2 \mathscr{P} \bigr)
	-
	N
	\mathscr{G}
	\biggl[
	- 2 \pi + \frac{\mathscr{P}}{3} \bigl( 2 \mathscr{K} \!-\! \mathscr{B} \bigr)
	\biggr]
		\Biggr\} \, .
\nonumber
\end{align}
Note that the symplectic part of this reduced action does not take the
canonical form, and consequently the brackets between canonical variables
are not canonical either. This is generally a consequence of reducing the
phase space by solving for the second-class constraints, that changes the Poisson
brackets to the generalized Dirac brackets. Nonetheless, here we can 
canonicalize the symplectic part by shifting the canonical momentum~$\pi^{ij}$,
accompanied by shifting the Lagrange multipliers that simplify the Hamiltonian
part of the action,
\begin{align}
\pi^{ij} \longrightarrow{}&
	\pi^{ij} 
	- \frac{\mathscr{M} {H^{ij}}_{kl} \pi^{kl} }{ \mathscr{M} \!+\! \mathscr{P} } 
	- \frac{ \mathscr{M} \mathscr{K} h^{ij}   }{9} 
	- \frac{ \mathscr{P} \mathscr{B} h^{ij}  }{9} 
	\, ,
\\
\mathscr{F} \longrightarrow{}&
	\mathscr{F} 
    +
    2 \mathscr{K} \mathscr{G}
	-
    \frac{ 9 \mathscr{M} \pi^{ij} H_{ijkl} \pi^{kl} }
        { \mathscr{P} \bigl( \mathscr{M} \!+\! \mathscr{P} \bigr)^2 }
	-
    \frac{ 2\mathscr{K} }{ 3\mathscr{P} } 
            \Bigl[ 6\pi 
            \!+\!
            \mathscr{K}
            \bigl( \mathscr{M} \!+\! 2\mathscr{P} \bigr)
            \Bigr]
\nonumber \\
&
    -
    \frac{ 6 H_{ijkl} \pi^{kl} }{ \bigl( \mathscr{M} \!+\! \mathscr{P} \bigr) } 
     \frac{ ( \nabla^i N^j ) }{N}
     -
      \frac{ 4 \mathscr{K} }{3}
        \frac{(\nabla^i N_i ) }{N}
    -
    \frac{2 \mathscr{K} N_i}{N} 
		\Bigl( \frac{ \mathscr{P} }{ \sqrt{h} } \Bigr)^{\!-1}
		\nabla^i \Bigl( \frac{ \mathscr{P} }{\sqrt{h}}  \Bigr)
	\, ,
\\
\mathscr{G} \longrightarrow{}&
	\mathscr{G} 
	+
	\frac{4}{3} \Bigl( 1
        \!+\! \frac{ \mathscr{M}  }{ \mathscr{P} }
         \Bigr) \mathscr{K}
    +
	\frac{2 \mathscr{B} }{3}
    +
	\frac{4(\nabla^i N_i )}{3N}
	+
 \frac{2N_i}{N} 
		\Bigl( \frac{ \mathscr{P} }{ \sqrt{h} } \Bigr)^{\!-1}
		\nabla^i \Bigl( \frac{ \mathscr{P} }{\sqrt{h}}  \Bigr)
  \, ,
\end{align}
upon which the reduced canonical action reads,
\begin{align}
\MoveEqLeft[0.5]
\mathscr{S}_{\rm red}
	\bigl[ h_{ij}, \pi^{ij} , \mathscr{K}, \mathscr{M} , \mathscr{B} , \mathscr{P} ,
	N, N_i,  \mathscr{F} , \mathscr{G} \bigr]
\nonumber \\
	&= \! \int\! d^{4\!}x \, \Biggl\{
		\pi^{ij}\partial_0 h_{ij}
		+ \frac{ \mathscr{M} \partial_0 \mathscr{K} }{3}
		+ \frac{ \mathscr{P} \partial_0 \mathscr{B} }{3}
	- N 
	\sqrt{h} \biggl[ 
	3 \frac{ \sqrt{h} }{ \mathscr{P} } 
		\biggl( \frac{ \pi^{ij} }{ \sqrt{h} } \frac{ \pi_{ij} }{ \sqrt{h} }
			- \frac{ \pi }{ \sqrt{h} } \frac{ \pi }{ \sqrt{h} } \biggr)
\nonumber \\
&
	-
	\frac{ \mathscr{P} }{ \sqrt{h} } \frac{ \mathcal{R} }{3} 
	+
	\frac{1}{36} \Bigl( \frac{ \mathscr{P} }{ \sqrt{h} } \Bigr)^{\!2}
	+ 
	\frac{2}{3} \nabla_i \nabla^i \Bigl( \frac{ \mathscr{P} }{ \sqrt{h} } \Bigr)	
	- 
	\frac{1}{2}\frac{\sqrt{h}}{ \mathscr{P} }
			\nabla_i \Bigl( \frac{ \mathscr{P} }{\sqrt{h}} \Bigr)
			\nabla^i \Bigl( \frac{ \mathscr{P} }{\sqrt{h}} \Bigr)
		\biggr] 
\nonumber \\
&
	- N_i
	\sqrt{h} \biggl[ 
	- 2 \nabla_j \Bigl( \frac{ \pi^{ij} }{ \sqrt{h} }  \Bigr)
	+
	2 \nabla^i \Bigl( \frac{\pi}{ \sqrt{h} } \Bigr)
	-
	2 \frac{\pi}{ \sqrt{h} } 
		\Bigl( \frac{ \mathscr{P} }{ \sqrt{h} } \Bigr)^{\!-1}
		\nabla^i \Bigl( \frac{ \mathscr{P} }{\sqrt{h}}  \Bigr)
	\biggr]	
\nonumber \\
&
	+ 
	\frac{N}{3} \mathscr{F} \bigl( \mathscr{M} \!+\! 2 \mathscr{P} \bigr)
	+
	N
	\mathscr{G}
	\biggl[
	2 \pi + \frac{\mathscr{P}}{3} \bigl( 2 \mathscr{K} \!-\! \mathscr{B} \bigr)
	\biggr]
		\Biggr\} \, .
\label{reduced canonical action}
\end{align}
This action might still look somewhat complicated, however it is in fact just a 
simple theory in disguise. This is revealed in two different ways that we consider 
below. The first is deriving the Lagrangian formulation of the reduced action that 
reveals it to be the Brans-Dicke formulation of metric-affine~$R^2$ 
gravity given in~(\ref{Brans-Dicke}). The second is trivialization of first-class 
constraints remaining from the affine connection sector, that reveals equivalence to 
the metric Einstein-Hilbert theory with a cosmological constant, that was derived in 
Sec.~\ref{sec: Equivalence to Einstein-Hilbert}. This also reveals how a dimensionful 
scale appears in the canonical formalism.

\medskip

\noindent {\bf Equivalence to conformal Brans-Dicke.} 
The Lagrangian formulation of the action~(\ref{reduced canonical action})
is obtained by first solving for
half of the canonical variables
and the two multipliers~$\mathscr{F}$
and~$\mathscr{G}$ in terms
of the first time derivatives and other 
variables.
This is accomplished by
varying the action above with respect to~$\pi^{ij}$,~$\mathscr{M}$,~$\mathscr{F}$,~$\mathscr{B}$, and~$\mathscr{G}$,
respectively,
\begin{align}
0 \approx{}&
	\partial_0 h_{ij}
	- \frac{6N}{ \mathscr{P} } \bigl( \pi_{ij} \!-\! h_{ij} \pi \bigr)
	- 2 \bigl( \delta_{(i}^k \delta_{j)}^l - h_{ij} k^{kl} \bigr) \nabla_{k} N_{l}
\nonumber \\
&	\hspace{3.5cm}
	+ 2 h_{ij} \Bigl( \frac{ \mathscr{P} }{ \sqrt{h} } \Bigr)^{\!-1}
		N^k \nabla_k \Bigl( \frac{ \mathscr{P} }{ \sqrt{h} } \Bigr)
	+ 2 N \mathscr{G} h_{ij}
	\, ,
\\
0 \approx{}&
	\partial_0 \mathscr{K} + N \mathscr{F} \, ,
\qquad\qquad
0 \approx
	\mathscr{M} + 2 \mathscr{P} \, ,
\\
0 \approx{}& 
	\partial_0 \mathscr{P} + N \mathscr{G} \mathscr{P} \, ,
\qquad\qquad
0 \approx 2 \pi + \frac{\mathscr{P}}{3} \bigl( 2 \mathscr{K} \!-\! \mathscr{B} \bigr) \, .
\end{align}
and solving for the said variables on-shell,
\begin{align}
\pi^{ij} \approx{}&
	- \frac{ \mathscr{P} K_{ij} }{ 3 } 
	+
	h_{ij} 
	\frac{ \sqrt{h} }{ 6 N }
	\biggl[
		\partial_0 \Bigl( \frac{ \mathscr{P} }{ \sqrt{h} } \Bigr)
		-
		N^k \nabla_k \Bigl( \frac{\mathscr{P}}{\sqrt{h}} \Bigr)
	\biggr]
	\, ,
\\
\mathscr{M} \approx{}&
	- 2 \mathscr{P} \, ,
\qquad
\mathscr{F} \approx 
	- \frac{\partial_0 \mathscr{K}}{ N } \, ,
\qquad
\mathscr{B} \approx
	2 \mathscr{K} + \frac{6\pi}{ \mathscr{P} }
	\, ,
\qquad
\mathscr{G} \approx - \frac{ \partial_0 \mathscr{P} }{N \mathscr{P}}
\, .
\end{align}
Here we have written time derivatives of the metric in terms of the extrinsic curvature defined
in~(\ref{extr curv def}). Plugging these solutions into the reduced canonical action~(\ref{reduced canonical action})
as off-shell equalities leads to the Lagrangian formulation of the action, from which the dependence on~$\mathscr{K}$
simply drops out,

\begin{equation}
S\bigl[ g_{\mu\nu} , \mathscr{P} \bigr]
    = \int\! d^{4\!}x \, \sqrt{-g} \, \biggl[
        \frac{ \mathscr{P} }{ \sqrt{h} }
            \frac{ \mathring{R} }{ 3 }
        + \frac{1}{2} \Bigl( \frac{\mathscr{P}}{\sqrt{h}} \Bigr)^{\!-1}
            \mathring{\nabla}^\mu 
                \Bigl( \frac{\mathscr{P}}{\sqrt{h}} \Bigr)
            \mathring{\nabla}_\mu
                \Bigl( \frac{\mathscr{P}}{\sqrt{h}} \Bigr)
        - \frac{1}{36} \Bigl( \frac{\mathscr{P}}{\sqrt{h}} \Bigr)^{\!2}
        \biggr]
        \, .
\end{equation}%
Upon recognizing a scalar field~$\mathscr{P}/\sqrt{h}\!\to\!3\sigma$
one obtains the Brans-Dicke formulation given in~(\ref{Brans-Dicke})
with~$\omega\!=\!-3/2$ and~$V(\sigma) \!=\! \sigma^2/4 $,
while defining a different scalar 
field~$\mathscr{P}/\sqrt{h} \!\to\! \phi^2/4$ 
puts the Lagrangian action into the form,
\begin{equation}
S_{\rm red}
	\bigl[ g_{\mu\nu}, \phi \bigr]
	= \int\! d^{4\!}x \,
	\sqrt{-g}
	\biggl[ 
	\frac{ \phi^2\mathring{R}}{12}
	+
	\frac{1}{2} \bigl( \mathring{\nabla}^\mu \phi \bigr) 
		\bigl( \mathring{\nabla}_\mu \phi \bigr)
	-
	\frac{\phi^4}{576}
	\biggr]
	\, ,
\end{equation}
that is often used in studies of scale invariant gravity (e.g.~\cite{Ferreira:2019ywk}).

\medskip

\noindent {\bf Trivializing Weyl symmetry in canonical formulation.}
It is known that if constraints can be promoted to 
canonical variables with canonical Poisson brackets
the pure gauge sector is guaranteed to decouple 
from the physical sector of the theory.
To this end we first adopt the
first-class constraints of the canonical 
action~(\ref{reduced canonical action}) as variables,
\begin{equation}
\mathscr{C} \equiv \frac{ \mathscr{M} \!+\! 2 \mathscr{P} }{3} \, ,
\qquad \qquad
\mathscr{X} \equiv 
	- \frac{ 2\pi }{ \mathscr{P} } - \frac{1}{3} \bigl( 2\mathscr{K} \!-\! \mathscr{B} \bigr)
	\, ,
\label{decoupling variables}
\end{equation}
instead of~$\mathscr{M}$ and~$\mathscr{B}$,
upon which the action takes the form,
\begin{align}
\MoveEqLeft[2]
\mathscr{S}_{\rm red}
	\bigl[ h_{ij}, \pi^{ij} , \mathscr{K}, \mathscr{C} , \mathscr{X} , \mathscr{P} ,
	N, N_i,  \mathscr{F} \bigr]
	= \int\! d^{4\!}x \, \Biggl\{
		\pi^{ij} \partial_0 h_{ij}
	+ 
	\mathscr{C} \partial_0 \mathscr{K}
	+ 
	\mathscr{P} \partial_0 \mathscr{X}
\nonumber \\
&
	-
	\frac{ 2 \pi }{\mathscr{P}} \partial_0 \mathscr{P}
	-
	N 
	\sqrt{h} \biggl[ 
	3 \frac{ \sqrt{h} }{ \mathscr{P} } 
		\biggl(
		\frac{ \pi^{ij} }{ \sqrt{h} } 
		\frac{ \pi_{ij} }{ \sqrt{h} }
		-
		\frac{ \pi }{ \sqrt{h} } 
		\frac{ \pi }{ \sqrt{h} }
		\biggr)
	-
	\frac{ \mathscr{P} }{ \sqrt{h} } \frac{ \mathcal{R} }{3} 
	+
	\frac{1}{36} \Bigl( \frac{ \mathscr{P} }{ \sqrt{h} } \Bigr)^{\!2}
\nonumber \\
&
	+ 
	\frac{2}{3} \nabla_i \nabla^i \Bigl( \frac{ \mathscr{P} }{ \sqrt{h} } \Bigr)
	- 
	\frac{1}{2}\frac{\sqrt{h}}{ \mathscr{P} }
			\nabla_i \Bigl( \frac{ \mathscr{P} }{\sqrt{h}} \Bigr)
			\nabla^i \Bigl( \frac{ \mathscr{P} }{\sqrt{h}} \Bigr)
		\biggr] 
	- 
	N_i
	\sqrt{h} \biggl[ 
	- 2
	\nabla_j \Bigl( \frac{ \pi^{ij} }{ \sqrt{h} }  \Bigr)
\nonumber \\
&
	+
	2 \nabla^i \Bigl( \frac{ \pi }{ \sqrt{h} } \Bigr)
	-
	2 \frac{ \pi }{ \sqrt{h} } \frac{ \sqrt{h} }{ \mathscr{P} } \nabla^i \Bigl( \frac{ \mathscr{P} }{ \sqrt{h} } \Bigr)
	\biggr]
	+ N \mathscr{F} \mathscr{C}
	- N \mathscr{G} \mathscr{X}
		\Biggr\} 
		\, .
\end{align}
Next the rescaling of the spatial metric and its canonical momentum,
\begin{equation}
h_{ij} \longrightarrow 
	\Bigl( c \frac{ \mathscr{P} }{ \sqrt{h} } \Bigr)^{\!2} h_{ij}
	\, ,
\qquad \qquad
\pi^{ij} \longrightarrow
	\Bigl( c \frac{ \mathscr{P} }{ \sqrt{h} } \Bigr)^{\!-2} 
	\Bigl( \pi^{ij} - \frac{ \pi h^{ij} }{2} \Bigr)
	\, ,
\label{rescale1}
\end{equation}
canonicalizes the symplectic part of the action.
Note that we required the metric to stay dimensionless,
and for that reason we had to introduce a dimensionful
scale~$c$ that is not present in the original formulation
of the theory. This is how a scale appears in the canonical
formulation --- by canonicalizing the symplectic part of the
action.
Following the rescaling in~(\ref{rescale1}) by a rescaling of lapse and shift,
\begin{equation}
N \longrightarrow
	\Bigl( c \frac{ \mathscr{P} }{ \sqrt{h} } \Bigr) N \, ,
\qquad \qquad
N_i \longrightarrow \Bigl( c \frac{ \mathscr{P} }{ \sqrt{h} } \Bigr)^{\!2} N_i \, ,
\end{equation}
and the remaining two Lagrange multipliers,
\begin{equation}
\mathscr{F} \longrightarrow 
	\Bigl( c \frac{ \mathscr{P} }{ \sqrt{h} } \Bigr)^{\!-1}
	\frac{\mathscr{F}}{N} 
	\, ,
\qquad \qquad
\mathscr{G} \longrightarrow
	\Bigl( c \frac{ \mathscr{P} }{ \sqrt{h} } \Bigr)^{\!-1}
	\frac{\mathscr{G}}{N} 
	\, ,
\end{equation}
accomplishes the task of decoupling the gauge sector 
associated to Weyl invariance from the physical sector,
\begin{align}
\MoveEqLeft[6]
\mathscr{S}_{\rm red}
	\bigl[ h_{ij}, \pi^{ij} , \mathscr{K}, \mathscr{C} , \mathscr{X} , \mathscr{P} ,
	N, N_i,  \mathscr{F}, \mathscr{G} \bigr]
\nonumber \\
&
 =
 \mathscr{S}_{\rm red}^{\rm phy}
	\bigl[ h_{ij}, \pi^{ij}, N , N_i \bigr]
 +
 \mathscr{S}_{\rm red}^{\rm gau}
	\bigl[ \mathscr{K}, \mathscr{C} , \mathscr{X} , \mathscr{P} ,
	\mathscr{F}, \mathscr{G} \bigr]
 \, .
\end{align}
The pure gauge part of the canonical action is,
\begin{equation}
 \mathscr{S}_{\rm red}^{\rm gau}
	\bigl[ \mathscr{K}, \mathscr{C} , \mathscr{X} , \mathscr{P} ,
	\mathscr{F}, \mathscr{G} \bigr]
 =
     \int\! d^{4}x
    \Bigl[
	\mathscr{C} \partial_0 \mathscr{K}
	+ 
	\mathscr{P} \partial_0 \mathscr{X}
    	+ \mathscr{F} \mathscr{C}
	- \mathscr{G} \mathscr{X}
		\Bigr]
  \, ,
\end{equation}
while the physical part is
the canonical formulation of Einstein-Hilbert action with 
a cosmological constant,
\begin{align}
 \mathscr{S}_{\rm red}^{\rm phy}
	\bigl[ h_{ij}, \pi^{ij}, N , N_i \bigr]
 ={}&
     \int\! d^{4\!}x \, \Biggl\{
	\pi^{ij} \partial_0 h_{ij}
	-
	N 
	\sqrt{h} 
	\biggl[ 
	3c \biggl(
		\frac{ \pi^{ij} }{ \sqrt{h} } \frac{ \pi_{ij} }{ \sqrt{h} } 
		-
		\frac{1}{2}
		\frac{ \pi }{ \sqrt{h} } \frac{ \pi }{ \sqrt{h} } 
		\biggr)
\nonumber \\
&
	-
	\frac{ 1 }{3c} \Bigl( \mathcal{R} - \frac{1}{12c} \Bigr)
	\biggr]
	-
	N_i
	\sqrt{h} 
	\biggl[ 
	- 2
	\nabla_j \Bigl( \frac{ \pi^{ij} }{ \sqrt{h} }  \Bigr)
	\biggr]
 \Biggr\}
 \, ,
\end{align}
where now we see the correspondence~$3c\!=\!\kappa^2$
between the dimensionful
scale introduced here and the dimensionful
scale introduced in Sec.~\ref{sec: Equivalence to Einstein-Hilbert} when deriving equivalence to the
Einstein-Hilbert action in~(\ref{EH action}).

\section{Discussion}
\label{sec: Discussion}

Motivated by the question of the number and character of degrees of freedom in general 
metric-affine theories, and the connection to projective- and Weyl-symmetry, in this 
work we worked out the technical details for the Hamiltonian constraint analysis 
necessary to address this question. It is particularly relevant to demonstrate that 
this approach works given the recent report claiming that the Dirac-Bergmann algorithm, 
that the Hamiltonian constraint analysis is based on, does not work for all field 
theories~\cite{DAmbrosio:2023asf}. We encounter no obstructions to this algorithm.
We first presented convenient ADM variables to use for the analysis of metric-affine 
gravity theories. As the independent connection can conveniently be represented as 
the Levi-Civita connection plus the distortion 3-tensor, we give the ADM decomposition
for a general 3-tensor field in Table~\ref{ADMtable1} and its first time derivatives 
in Table~\ref{ADMtable2}. For theories with projective invariance it is appropriate to
split the distortion tensor into non-metricity and the 
projective-invariant~$B$-tensor~(\ref{B tensor}). The ADM decomposition for both tensors 
and their time derivatives is given in 
Tables~\ref{ADMnonmetricityTable}--\ref{ADM Btensor ders}. We have demonstrated 
the utility of these variables by performing the Hamiltonian constraint analysis of the 
metric-affine-$R^2$ theory in a way that mirrors as much as possible the steps one would 
have to go through when analyzing more general metric-affine theories of gravity. This 
is a highly informative example as results of the analysis can readily be compared to 
properties of the model known in the literature. Furthermore, the analysis of this model 
demonstrates powerful properties of the Hamiltonian analysis:  the casting of the theory 
into a far simpler form by the solving of second-class constraints (which reveals the 
theory to have the same canonical form as a specific Brans-Dicke theory), and the 
trivialization of symmetry (which reveals the theory to have the same canonical form as 
General Relativity in the presence of a cosmological constant). It is conceivable the 
Hamiltonian analysis of the wider set of Weyl- and projective-invariant metric-affine 
theories of gravity will allow for the discovery of a more general correspondence of 
metric-affine theories to Riemannian theories of gravity coupled to specific matter 
content.

Understanding the physical content of a pure modified gravity theory, such as the 
metric-affine theories we consider here, is not by itself sufficient to decide on the 
viability of the theory. Ultimately, this is decided by considering how matter couples to
a given modified gravity theory. This is what should be explored for the general metric 
affine, projective-invariant, and Weyl-invariant theories that we set out to study, after 
the pure gravity sector has been fully understood. It is generally not a straightforward 
task to understand the consequences of different ways of coupling matter to gravity. 
Before constraining the particular modified gravity model and its coupling to matter it is 
paramount to understand theoretical aspects of the theory. The primary question of 
interest is the number of propagating degrees of freedom in the given theory. The prime 
example of how nontrivial this question is are purely metric scalar-tensor theories. 
Requiring that such theories propagate a single scalar and a single graviton severely 
restricts the forms of coupling to those of Horndeski theories~\cite{Horndeski:1974wa} 
(and their generalizations so-called beyond-Horndeski~\cite{Gleyzes:2014dya} 
and DHOST theories~\cite{Langlois:2015cwa}). Such constraints on couplings are also known 
for $U(1)$ vector-tensor theories~\cite{Horndeski:1976gi}. Contrary to pure metric 
theories, coupling matter to metric-affine gravity is comparatively far less explored and 
understood. Part of the reason behind this is the much greater number of possibilities due 
to the additional independent connection structure. Nonetheless, it is still possible to 
use simple methods of conformal rescaling from 
Sec.~\ref{sec: Equivalence to Einstein-Hilbert} to analyze theories where matter couples 
to the metric and the metric-affine Ricci scalar. A more sophisticated method 
of field redefinitions can be used to analyze the so-called Ricci-based 
gravity~\cite{BeltranJimenez:2019acz,Afonso:2017bxr,Borowiec:1996kg}, the metric-affine 
theory of gravity where matter couples to the metric and the symmetric part of the 
metric-affine Ricci tensor~$R_{(\mu\nu)}$. However, in more general situations where 
other Riemann tensor contractions appear either in the pure gravity sector or 
in the matter coupling, one is typically limited to examining linearized perturbations
around particular backgrounds~\cite{Annala:2022gtl}. While such studies are insightful 
and informative, the ultimate answer on the number of degrees of freedom should be decided 
by the full Hamiltonian analysis machinery, such as the one we outlined here. Restricting 
ourselves to projective- and Weyl-invariant couplings it is straightforward to write down 
an example of the scalar coupling to curvature
coupling --- $\bigl( a R^{(\mu\nu)} + b \hat{R}^{(\mu\nu)} \bigr) \partial_\mu \Phi \partial_\nu \Phi$, with~$\Phi$ the scalar field of conformal weight zero
---
for which it is not known how it influences the number of propagating degrees of freedom.
While couplings such as these are classified in detail for metric 
scalar-tensor theories, investigation of this matter in metric-affine 
scalar-tensor theories is in its early stages (see 
e.g.~\cite{Galtsov:2018xuc,Helpin:2019vrv,Helpin:2019kcq,Aoki:2019rvi,Ikeda:2023dzr,Bahamonde:2022cmz}).

\section*{Acknowledgements}

We thank William Barker, Jose Beltr\'{a}n Jim\'{e}nez, and Tomi Koivisto for helpful discussions. DG was supported by the European Union and the Czech Ministry of Education, 
Youth and Sports 
(Project: MSCA Fellowship CZ FZU I --- 
\\
CZ.02.01.01/00/22\textunderscore010/0002906).
TZ acknowledges support from the project No. 2021/43/P/ST2/02141 co-funded by 
the Polish National Science Centre and the European Union Framework Programme 
for Research and Innovation Horizon 2020 under the Marie Sk\l{}odowska-Curie grant 
agreement No. 945339.
The work of CL was supported by the grants
2018/30/Q/ST9/00795 and 2021/42/E/ST9/00260 from
the Polish National Science Centre.



\end{document}